\documentclass[12pt,a4paper]{article}
\pdfoutput=1
\usepackage{bm, amssymb,pifont,cancel, amsmath,comment,color}
\usepackage{here}
\usepackage{cite}
\usepackage{graphicx}
\usepackage{subfigure}

\makeatletter

\@addtoreset{equation}{section}
\makeatother
 
\begin{document}

\begin{titlepage}
\null
\begin{flushright}
WU-HEP-20-15
\end{flushright}

\vskip 1cm
\begin{center}
\baselineskip 0.8cm
{\LARGE \bf
Disentangling mass spectra of multiple fields\\ in cosmological collider
}

\lineskip .75em
\vskip 1cm

\normalsize

{\large } {\large Shuntaro Aoki} $^{1}${\def\thefootnote{\fnsymbol{footnote}}\footnote[1]{E-mail address: shun-soccer@akane.waseda.jp}},
{\large Masahide Yamaguchi} $^{2}${\def\thefootnote{\fnsymbol{footnote}}\footnote[2]{E-mail address: gucci@phys.titech.ac.jp}},

\vskip 1.0em

$^1${\small\it Department of Physics, Waseda University, \\ 
Tokyo 169-8555, Japan}
\vskip 1.0em

$^2${\small{\it Department of Physics, Tokyo Institute of Technology,\\ Tokyo 152-8551, Japan}}
\vskip 1.0em

\vspace{12mm}

{\bf Abstract}\\[5mm]
{\parbox{13cm}{\hspace{5mm} \small
We study effects of multiple scalar fields (scalar isocurvatons) with the Hubble scale masses on the inflationary bispectrum in the squeezed limit, particular paying attention to the question how to disentangle mass spectra of such fields. We consider two isocurvatons with almost degenerate masses and the coupling of an inflaton to both isocurvatons as an example. We find that the characteristic feature associated with nearly degenerate masses appears in the oscillating part of the bispectrum, which is dominated by a waveform with a specific wavelength roughly given by an inverse of the mass difference. Such a waveform with a relatively longer wavelength can be easily identified and useful for disentangling almost degenerate mass spectra. This situation is in sharp contrast with the case of collider experiments on earth, where the very precise energy resolution corresponding to the mass difference is required to disentangle almost degenerate mass spectra. 
Therefore, if future observations could detect this kind of a characteristic feature in bispectrum of the primordial curvature perturbations, it can prove the existence of degenerate multiple particles around the Hubble scale and resolve their mass degeneracies.

}}

\end{center}

\end{titlepage}

\tableofcontents
\vspace{35pt}
\hrule

\section{Introduction}\label{intro}
It has been recently claimed that cosmological higher order correlation functions of the primordial curvature perturbations play important roles not only for restricting inflation models, but for exploring a new massive particle (isocurvaton) with mass up to the Hubble scale~$H$. In particular, when there exist couplings between an inflaton and an isocurvaton, we expect an oscillatory shape in the squeezed limit of the three point correlation (bispectrum), which contains information of mass as well as spin of the isocurvaton. This approach known as the cosmological collider has attracted much attention so far~\cite{Chen:2009we,Chen:2009zp,Baumann:2011nk,Assassi:2012zq,Sefusatti:2012ye,Norena:2012yi,Chen:2012ge,Pi:2012gf,Noumi:2012vr,Cespedes:2013rda,Gong:2013sma,Emami:2013lma,Kehagias:2015jha,Liu:2015tza,Arkani-Hamed:2015bza,Dimastrogiovanni:2015pla,Schmidt:2015xka,Chen:2015lza,Delacretaz:2015edn,Bonga:2015urq,Flauger:2016idt,Lee:2016vti,Delacretaz:2016nhw,Meerburg:2016zdz,Chen:2016uwp,Chen:2016hrz,Kehagias:2017cym,An:2017hlx,Tong:2017iat,Iyer:2017qzw,An:2017rwo,Kumar:2017ecc,Riquelme:2017bxt,Franciolini:2017ktv,Saito:2018omt,Cabass:2018roz,Wang:2018tbf,Chen:2018xck,Bartolo:2018hjc,Dimastrogiovanni:2018uqy,Bordin:2018pca,Achucarro:2018ngj,Chua:2018dqh,Arkani-Hamed:2018kmz,Kumar:2018jxz,Goon:2018fyu,Wu:2018lmx,Anninos:2019nib,Li:2019ves,McAneny:2019epy,Kim:2019wjo,Lu:2019tjj,Hook:2019zxa,Hook:2019vcn,Kumar:2019ebj,Liu:2019fag,Wang:2019gbi,Wang:2020uic,Li:2020xwr,Wang:2020ioa,Bodas:2020yho}, because the energy scale that can be reached in ground experiments has almost reached its limit.  

As far as we aware of it, almost all of the works done thus far concentrate on understanding the signal of an individual particle with a definite spin. However, in the realistic situations, there exist more than one particle which couple to an inflaton, and their masses lie around the Hubble scale. In Ref.~\cite{Chen:2016nrs}, it is shown that some of the standard model particles acquire masses of about $H$ during inflation due to the loop effects. Their effects are classified in Refs.~\cite{Chen:2016uwp,Chen:2016hrz}, in order to distinguish a new physics from the standard model background. Moreover, in supersymmetric inflation scenarios, several scalar fields with couplings to an inflaton naturally have tree level masses around the Hubble scale (Hubble induced mass) by supergravity corrections~\cite{Stewart:1994ts}. Then, the following question arises; how to disentangle degenerate mass spectra of multiple particles in the cosmological collider. In the case of collider 
 experiments on earth, the very precise energy resolution corresponding to the mass difference is required to resolve such a degeneracy. One may wonder if we also need the precise measurement corresponding to the mass difference to disentangle degenerate mass spectra of multiple particles in the cosmological collider. This is the main topic we are going to address in this paper.

For this purpose, in this paper, we discuss the effect of multiple isocurvatons on the bispectrum, focusing on the oscillatory behavior in the squeezed limit. We consider a case where the isocurvatons are scalar fields as an example. As we will see, the total signal becomes a superposition of different waveforms, and as a result, some characteristic oscillation behavior appears. Also, we will carefully analyze the mixing effect between two isocurvatons, which leads to a specific signal on the bispectrum. 

The paper is organized as follows. In Sec.~\ref{setup}, we prepare some necessary equations and specify the interactions. We assume simple non-derivative couplings between an inflaton and isocurvatons. Based on them, we give an analytic formula of the bispectrum with multiple isocurvatons under the super-horizon approximation in Sec.~\ref{bispectrum}. Then, we will study the signal by taking two fields case as an example, and discuss its observational consequence, though the extension to the case with more fields is straightforward. Section~\ref{summary} is devoted to the summary. We summarize some technicalities of the calculations in Appendix~\ref{perturbation} and~\ref{derivation}. The results of other cases with different inflaton-isocurvatons interactions are shown in Appendix~\ref{shift}. In Appendix~\ref{Origin}, we discuss an origin and another interpretation of our results, by moving to different bases of the field space.

Throughout the paper, we use the unit $M_{P}= 1$, where $M_{P}= 2.4\times 10^{18}$ GeV is the reduced Planck mass.

\section{Setup}\label{setup}
In this section, we specify our setup and derive the perturbed action around the inflationary background to compute the bispectrum in the next section.

\subsection{Action and background equations}
We consider the following action:
\begin{align}
\nonumber S=&\int d^4x \sqrt{-g}\biggl[\frac{1}{2}R-\frac{1}{2}g^{\mu\nu}\partial_{\mu}\phi\partial_{\nu}\phi-V(\phi)\\
&-\frac{1}{2}g^{\mu\nu}\delta_{IJ}\partial_{\mu}\sigma^I\partial_{\nu}\sigma^J-U(\sigma)\biggr]+S_{\rm{int}}(\phi,\sigma),\label{action}
\end{align}
where $\phi$ and $\sigma^I (I=1,\cdots, n)$ are an inflaton and isocurvatons, respectively. The last term, $S_{\rm{int}}$, describes the direct interaction between $\phi$ and $\sigma^I$, which will be specified later.\footnote{We could introduce more general interactions between the inflaton and the isocurvatons. A resultant perturbed action in that case can be found, e.g., in Ref.~\cite{Pinol:2020kvw}.}

The background fields satisfy 
\begin{align}
&\ddot{\phi}_0+3H\dot{\phi}_0+V_{\phi}=0,\\
&3H^2=\frac{1}{2}\dot{\phi}_0^2+V,\\
&U_I=0,
\end{align}
where the indices~$\phi$ and $I$ denote the derivative w.r.t $\phi$ and $\sigma^I$, and the subscript $``0"$ is attached on the background field. The dot denotes the time derivative. We assume that $\sigma^I$ takes a trivial vacuum expectation value during inflation,~$\sigma_0^I=0$, and $U=0$ at the background. 
\subsection{Quadratic action and mode functions}
To compute the bispectrum by a perturbative approach, we expand all of the fields and the metric around the background. Then, as usual, we quantize the free fields determined by the quadratic action, and treat the remaining parts as small perturbations.

Taking the so-called flat gauge, and solving Hamilton and momentum constraints (see Appendix~\ref{perturbation_free} for detail), we obtain the quadratic part of the action,
\begin{align}
\nonumber S^{(2)}=&\int d^4x a^3\biggl[ \frac{1}{2}(\delta \dot{\phi})^2-\frac{1}{2a^2}(\partial \delta \phi)^2-\frac{1}{2}\left(V_{\phi\phi}-2H^2(3\epsilon-\epsilon^2+\epsilon\eta)\right)(\delta\phi)^2\\
&+\frac{1}{2}( \dot{\sigma}^I)^2-\frac{1}{2a^2}(\partial  \sigma^I)^2-\frac{1}{2}U_{IJ}\sigma^I\sigma^J\biggr] ,\label{free2}
\end{align}
where we expand the fields as $\phi=\phi_0+\delta \phi$ and $\sigma^I=0+\delta \sigma^I$, and rewrite $\delta\sigma^I\rightarrow \sigma^I$. The derivatives~$\partial$ appearing above are the spatial ones. Now, we consider the case with $U_{IJ}={\rm{diag}}((m^1)^2,\cdots,(m^n)^2)$, which can be taken without loss of generality.

We quantize the fluctuations $\delta \phi_{\bm{k}}$ and $\sigma^I_{\bm{k}}$ in Fourier space as\footnote{We use a bold character such as ${\bm{k}}$ to denote three dimensional vectors and $k\equiv |{\bm{k}}|$ for their absolute values.} 
\begin{align}
&\delta \phi_{\bm{k}} =u_{k}a_{\bm{k}}+u_{k}^*a^{\dagger}_{-\bm{k}},\\
&\sigma^I_{\bm{k}} =v_{k}^Ib_{\bm{k}}^I+v_{k}^{I*}b^{I\dagger}_{-\bm{k}},
\end{align}
where the annihilation and the creation operators, $a_{\bm{k}},a^{\dagger}_{\bm{k}'},b_{\bm{k}}^I, b^{J\dagger}_{\bm{k}'}$,
satisfy the following commutation relations,
\begin{align}
[a_{\bm{k}},a^{\dagger}_{\bm{k}'}]=(2\pi)^3\delta^{(3)}({\bm{k}}-{\bm{k}'}), \ \ [b_{\bm{k}}^I,b^{J\dagger}_{\bm{k}'}]=(2\pi)^3\delta^{IJ}\delta^{(3)}({\bm{k}}-{\bm{k}'}).
\end{align}
The mode functions $u_k$ and $v_k^I$ satisfy the equations of motions from Eq.~$\eqref{free2}$,
\begin{align}
&(au_{k})''+\left(k^2-\frac{2}{\tau^2}\right)(au_{k})=0,\\
&(av^{I}_{k})''+\left(k^2-\frac{2}{\tau^2}+\frac{(m^I)^2}{H^2\tau^2}\right)(av_{k}^{I})=0,
\end{align}
where the prime denotes the derivative with respect to the conformal time $\tau\equiv \int a^{-1} dt$. In the derivation, we have neglected the mass term of $\delta \phi$, which is small during a slow-roll inflation. 
Their solutions are given by
\begin{align}
u_k=\frac{H}{\sqrt{2k^3}}(1+ik\tau)e^{-ik\tau},
\end{align}
and 
\begin{align}
v_k^I=-ie^{i\left(\nu^I +\frac{1}{2}\right)\frac{\pi}{2}}\frac{\sqrt{\pi}}{2}H(-\tau)^{3/2}H_{\nu^I}^{(1)}(-k\tau),
\end{align}
where $H_{\nu}^{(1)}$ is Hankel function of the first kind. The parameters $\nu^I$ are determined by the masses of the isocurvatons as
\begin{align}
\nu^I \equiv \sqrt{\frac{9}{4}-\left(\frac{m^I}{H}\right)^2},
\end{align}
which is either real or pure imaginary.\footnote{To avoid confusion, we use a subscript instead of a superscript for the mass labels in the following.}


\subsection{Interactions}
Here we introduce interactions between an inflaton and isocurvatons. Following Refs.~\cite{Chen:2016uwp,Chen:2016hrz}, we assume the form of the interactions as 
\begin{align}
S_{\rm{int}}(\phi,\sigma)=\int d^4x \mathcal{L}_{\rm{int}}=-\int d^{4} x \sqrt{-g}f(\phi,X)  g_{IJ}\sigma^I\sigma^J, \label{int_f} 
\end{align}
where $X\equiv g^{\mu\nu}\partial_{\mu}\phi\partial_{\nu}\phi$, and $g_{IJ}$ with $I,J=1,\cdots,n$ are real and symmetric constants. 
Expanding Eq.~$\eqref{int_f}$ around the background, we find that it contains three and four point vertices such as
\begin{align}
&\mathcal{L}^3_{\rm{int}}= -a^3f_{\phi}\delta \phi g_{IJ}\sigma^I\sigma^J,\\
&\mathcal{L}^4_{\rm{int}}=-\frac{1}{2}a^3f_{\phi\phi}(\delta \phi)^2 g_{IJ}\sigma^I\sigma^J,
\end{align}
which we are interested in this paper. More generally, since the three and the four point vertices can come mainly from different interactions, e.g., Eq.~$\eqref{int_f}$ and another one $\tilde{f}(\phi,X)  \tilde{g}_{IJ}\sigma^I\sigma^J$, we parametrize them as 
\begin{align}
&\mathcal{L}^3_{\rm{int}}=a^3c_3\delta \phi g_{IJ}\sigma^I\sigma^J,\label{L_3}\\
&\mathcal{L}^4_{\rm{int}}=a^3c_4(\delta \phi)^2 \tilde{g}_{IJ}\sigma^I\sigma^J,\label{L_4}
\end{align}
where $c_3\equiv -f_{\phi}$ and $c_4\equiv -\tilde{f}_{\phi\phi}/2$ are in general time dependent coupling constants, but are treated as constants under the de Sitter approximation. Note that Eq.~$\eqref{int_f}$ leads to different type of vertices unlike Eqs.~$\eqref{L_3}$ and~$\eqref{L_4}$, which are shown in Appendix~\ref{perturbation_int}.

\begin{figure}[t]
\centering
\includegraphics[width=8.0cm]{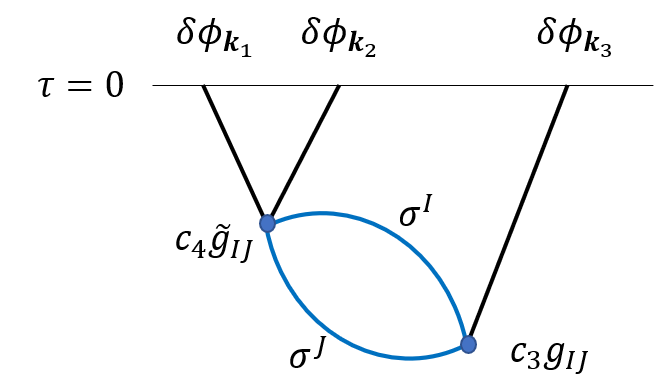}
\caption{Contribution of isocurvatons~$\sigma^I$ to the bispectrum}
\label{fig_diagram}
\end{figure}

Based on the interactions given above, we can consider the diagram shown in Fig.~\ref{fig_diagram}, which is a target of the following section.

\section{Bispectrum}\label{bispectrum}

In this section, we show the contribution of the diagram (Fig.~\ref{fig_diagram}) to the bispectrum.\footnote{The same diagrams with replacement of ${\bm{k}}_3\leftrightarrow {\bm{k}}_{1,2}$ do not lead to the so-called nonlocal momentum dependence which produces an oscillating signal in the squeezed limit $(k_1\sim k_2 \gg k_3)$. Hence, we do not consider them.}
The result of the bispectrum in the squeezed limit $(k_1\sim k_2 \gg k_3)$ at $\tau=0$ is given by
\begin{align}
\nonumber &\left\langle\delta \phi_{\bm{k}_{1}} \delta \phi_{\bm{k}_{2}} \delta \phi_{\bm{k}_3}\right\rangle '  \\
\nonumber =&\frac{H^2c_3c_4}{4 \pi^4}\sum_{I,J}g_{IJ}\tilde{g}_{IJ}\frac{1}{k_1^6}\biggl\{ C(\nu^I,\nu^J)\left(\frac{k_3}{2k_1}\right)^{-(\nu^I+\nu^J)}+(\nu^I\rightarrow \nu^I,\nu^J\rightarrow -\nu^J)\\
&+(\nu^I\rightarrow -\nu^I,\nu^J\rightarrow \nu^J)+(\nu^I\rightarrow -\nu^I,\nu^J\rightarrow -\nu^J)\biggr\}, \label{result}
\end{align}
where
\begin{align}
\nonumber C(\nu^I,\nu^J)=&(1-\nu^I-\nu^J)^2(4-\nu^I-\nu^J)\Gamma^2(-\nu^I-\nu^J)\Gamma(\nu^I)\Gamma(\nu^J)\\
\nonumber &\times \Gamma(\frac{3}{2}-\nu^I)\Gamma(\frac{3}{2}-\nu^J)\Gamma(-4+2(\nu^I+\nu^J))\\
&\times {\rm{cos}}(\frac{\pi}{2}(\nu^I+\nu^J)){\rm{sin}}^3(\frac{\pi}{2}(\nu^I+\nu^J)).\label{C}
\end{align}
The prime of the left-hand-side equation implies the omission of the delta function coming from the total momentum conservation,~$(2\pi)^3\delta^{(3)} ({\bm{k}_{1}}+{\bm{k}_{2}}+{\bm{k}_{3}})$. Note that the function~$C(\nu^I,\nu^J)$ is symmetric under the interchange of $\nu^I$ and $\nu^J$. 
We leave the detailed derivation of Eq.~$\eqref{result}$ to Appendix~\ref{derivation}.

When $I=J$ inside the summation, the second and the third terms in the right hand side of Eq.~$\eqref{result}$ do not have the momentum dependence, and the coefficients~$C$ in this case are evaluated as 
\begin{align}
C(\nu^I,-\nu^I) =C(-\nu^I,\nu^I) = \frac{\pi^3}{96}\Gamma(\nu^I)\Gamma(-\nu^I)\Gamma(\frac{3}{2}+\nu^I)\Gamma(\frac{3}{2}-\nu^I), \label{C_limit}  
\end{align}
by taking a limit $\nu^I=\nu^J$ carefully. 
 
Finally, by recalling the relation in the flat gauge between the inflaton fluctuation $\delta \phi$ and the curvature perturbation $\zeta$, $\zeta = -\frac{H}{\dot{\phi}_0}\delta \phi$, the bispectrum of the curvature perturbations can be obtained through Eq.~$\eqref{result}$ as
\begin{align}
\left\langle \zeta_{\bm{k}_{1}} \zeta_{\bm{k}_{2}} \zeta_{\bm{k}_3}\right\rangle '=(2\pi)^4P_{\zeta}^2\frac{1}{(k_1k_2k_3)^2}S(k_1,k_2,k_3),    
\end{align}
where $P_{\zeta}=\frac{H^4}{4\pi^2\dot{\phi}_0^2}$ is the power spectrum of $\zeta$ and $S$ is a shape function. In the squeezed limit of the bispectrum of the curvature perturbations, the shape function $S$ takes the following form, 
\begin{align}
 \nonumber S=&-\frac{c_4}{2P_{\zeta}^{1/2}\pi^5}\frac{c_3}{H}\sum_{I,J}g_{IJ}\tilde{g}_{IJ}\biggl\{ C(\nu^I,\nu^J)\left(\frac{k_3}{2k_1}\right)^{2-(\nu^I+\nu^J)}+(\nu^I\rightarrow \nu^I,\nu^J\rightarrow -\nu^J)\\
&+(\nu^I\rightarrow -\nu^I,\nu^J\rightarrow \nu^J)+(\nu^I\rightarrow -\nu^I,\nu^J\rightarrow -\nu^J)\biggr\}.
\label{shape}
\end{align}
Note that $g_{IJ},\tilde{g}_{IJ}$ and $c_4$ are dimensionless constants and that $c_3/H$ is a dimensionless combination. In the next section, we will investigate the multiple field effects based on the shape function~$\eqref{shape}$, instead of Eq.~$\eqref{result}$.

\section{Two fields example}\label{example}

To see the effects of multiple isocurvatons, let us consider the two fields case, $n=2$, as an example. The extension to the case with more fields is straightforward. Here, we consider the case where the masses of two isocurvatons satisfy $m_{1}, m_{2}>\frac{3}{2}H$.

Then, the total bispectrum is given by
\begin{align}
S=S_{11}+S_{22}+S_{12},
\end{align}
where 
\begin{align}
\nonumber S_{11}=&-\frac{c_4}{P_{\zeta}^{1/2}\pi^5}\frac{c_3}{H}g_{11}\tilde{g}_{11}\left(\frac{k_3}{2k_1}\right)^2\biggl[ {\rm{Re}}C(i\mu^1,i\mu^1){\rm{cos}}\left(2\mu^1\log\left(\frac{k_3}{2k_1} \right)\right)\label{case1S_11}\\
&+{\rm{Im}}C(i\mu^1,i\mu^1){\rm{sin}}\left(2\mu^1\log\left(\frac{k_3}{2k_1} \right)\right) +C(i\mu^1,-i\mu^1)\biggr],\\
S_{22}=&\mu^1\rightarrow \mu^2\ {\rm{and}}\ g_{11}\tilde{g}_{11}\rightarrow g_{22}\tilde{g}_{22}\ {\rm{in}}\ S_{11},\label{case1S_22}\\
\nonumber S_{12}=&-\frac{2c_4}{P_{\zeta}^{1/2}\pi^5}\frac{c_3}{H}g_{12}\tilde{g}_{12}\left(\frac{k_3}{2k_1}\right)^2\biggl[ {\rm{Re}}C(i\mu^1,i\mu^2){\rm{cos}}\left((\mu^1+\mu^2)\log\left(\frac{k_3}{2k_1} \right)\right)\\
\nonumber &+{\rm{Im}}C(i\mu^1,i\mu^2){\rm{sin}}\left((\mu^1+\mu^2)\log\left(\frac{k_3}{2k_1} \right)\right) \\
\nonumber &+{\rm{Re}}C(i\mu^1,-i\mu^2){\rm{cos}}\left((\mu^1-\mu^2)\log\left(\frac{k_3}{2k_1} \right)\right)\\
&+{\rm{Im}}C(i\mu^1,-i\mu^2){\rm{sin}}\left((\mu^1-\mu^2)\log\left(\frac{k_3}{2k_1} \right)\right)\biggr].\label{case1S_12}
\end{align}
We introduced a notation, $\nu^I=i\mu^I$ with $\mu^I$ real.

\subsection{Special cases}
Before going to see the detailed analysis of the above equations, let us discuss their some special limits. 

\begin{itemize}
 \item \subsubsection*{Single field limit}

When $\mu^2\gg \mu^1$, the heavier field~$\sigma_2$ effectively decouples from the system, and hence we should recover the result of a single isocurvaton case~\cite{Chen:2016uwp}. Based on the Stirling expansion,
\begin{align}
|\Gamma(a+ib)|\sim \sqrt{2\pi}|b|^{a-1/2}e^{-\pi |b|/2},\ \ |b|\gg 1,\ \ a,b \in\mathbb{R},\label{Stirling}
\end{align}
we find that 
$S_{22}$ and $S_{12}$ in Eqs.~$\eqref{case1S_12}$ and $\eqref{case1S_22}$ are suppressed by 
\begin{align}
C(i\mu^2,\pm i\mu^2)\propto e^{-2\pi\mu^2},  \ \ 
C(i\mu^1,\pm i\mu^2)\propto e^{-\pi\mu^2}, \label{C_pm_asymp}
\end{align}
for $\mu^2\rightarrow \infty$ with $\mu^1$ fixed.
Therefore, we reproduce the prediction of the single isocurvaton case in this limit as it should be.

\item\subsubsection*{$C(i\mu^I,i\mu^J)$ vs. $C(i\mu^I,-i\mu^J)$ in degenerate limit}
Here, we examine the behavior of the coefficients $C(i\mu^I,i\mu^J)$ and $C(i\mu^I,-i\mu^J)$ for $I,J=1,2$, in the degenerate limit $\mu^1\sim \mu^2 \equiv \mu\gg 1$.

As regards $C(i\mu^I,i\mu^J)$, by using Eq.~$\eqref{Stirling}$, we can approximate it as 
\begin{align}
|C(i\mu^I,i\mu^J)|\sim  \frac{\pi^{7/2}}{2^7\sqrt{2}}\mu^{-3/2}e^{-2\pi \mu}, \label{Cplus_asymp}
\end{align}
for $I,J=1,2$. Here we included the prefactor in Stirling formula.  
On the other hand, for $C(i\mu^I,-i\mu^J)$, we obtain  
\begin{align}
 |C(i\mu^I,-i\mu^J)| \sim \frac{\pi^{5}}{4!} \mu e^{-2\pi \mu}, \label{Cminus_asymp}
\end{align}
in the same limit.

As one can see, the coefficients $C(i\mu^I,i\mu^J)$ and $C(i\mu^I,-i\mu^J)$ decay by Boltzmann suppression in a similar way, but their prefactors have different mass dependence. In particular, we have a relation 
\begin{align}
|C(i\mu^I,i\mu^J)/C(i\mu^I,-i\mu^J)|\sim 2\times 10^{-2} \times\mu^{-5/2},\ \ {\rm{for}}\ \  \mu\gg 1, \label{c+vsc-}
\end{align}
which plays an important role later when we discuss waveforms of the signal. 
\item\subsubsection*{$\mu^I=0$ for $I=1,2$}
Finally, let us comment on the limit $\mu^1=0$ and/or $\mu^2=0$. Indeed, the coefficient $C$ diverges in this limit, but the bispectrum is still finite. For example, taking $\mu^1\rightarrow 0$ in $S_{11}$, we obtain
\begin{align}
S_{11}\rightarrow -\frac{c_4}{2P_{\zeta}^{1/2}\pi^5}\frac{c_3}{H}g_{11}\tilde{g}_{11}\left(\frac{k_3}{2k_1}\right)^2\times \frac{\pi^4}{96}\left(\log \left(\frac{k_3}{2k_1}\right)\right)^2.
\end{align}
In the same way, $S_{12}$ and $S_{22}$ are also finite in this limit.
\end{itemize}

\subsection{Without mixing}\label{Without mixing}
Now, let us discuss the effects of multiple isocurvatons,  starting from the case without the mixing terms, i.e., $g_{12}=0$ or $\tilde{g}_{12}=0$. 
In this case, the bispectrum becomes a summation of two independent oscillating modes, $S_{11}$ and $S_{22}$. As a general structure of the oscillating signals, their amplitude and wavelength become smaller as $\mu^I$ becomes larger. In particular, as we saw in the previous subsection, the amplitude quickly decays for large $\mu^I$ by the Boltzmann suppression factor in $C$. In the following discussion, we set $c_4=c_3/H=1$ for simplicity and take $P_{\zeta}=2\times 10^{-9}$ to fit the observational data.\footnote{We use these values throughout the paper. Also, we require that the effective couplings satisfy $\frac{c_3}{H}e^{-\pi(\mu^I+\mu^J)}g_{IJ}<1$ and $c_4e^{-\pi(\mu^I+\mu^J)}\tilde{g}_{IJ}<1$ for the perturbation to be valid.}

\begin{figure}[H]
 \begin{minipage}{0.5\hsize}
  \begin{center}
   \includegraphics[width=70mm]{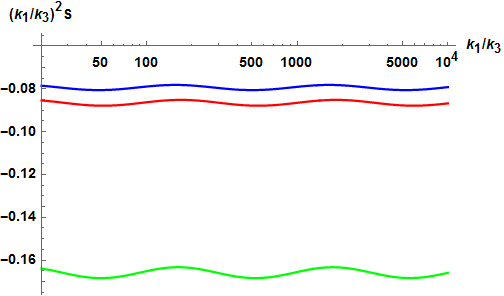}
  \end{center}
 \end{minipage}
 \begin{minipage}{0.5\hsize}
  \begin{center}
   \includegraphics[width=70mm]{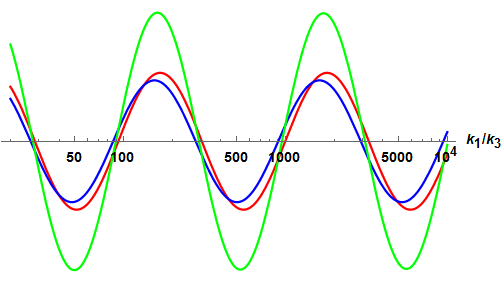}
  \end{center}
 \end{minipage}
 \begin{minipage}{0.5\hsize}
  \begin{center}
   \includegraphics[width=70mm]{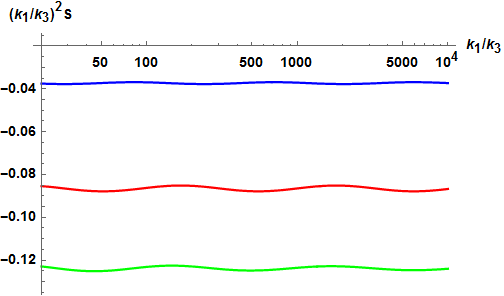}
  \end{center}
 \end{minipage}
 \begin{minipage}{0.5\hsize}
  \begin{center}
   \includegraphics[width=70mm]{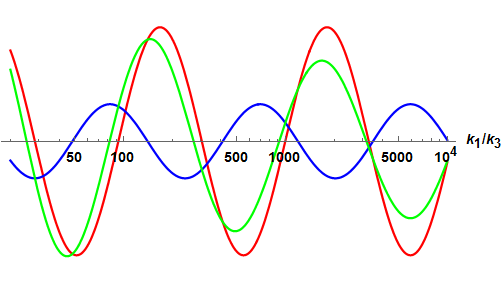}
  \end{center}
 \end{minipage}
  \begin{minipage}{0.5\hsize}
  \begin{center}
   \includegraphics[width=70mm]{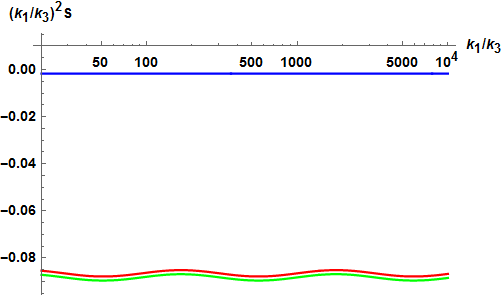}
  \end{center}
 \end{minipage}
 \begin{minipage}{0.5\hsize}
  \begin{center}
   \includegraphics[width=70mm]{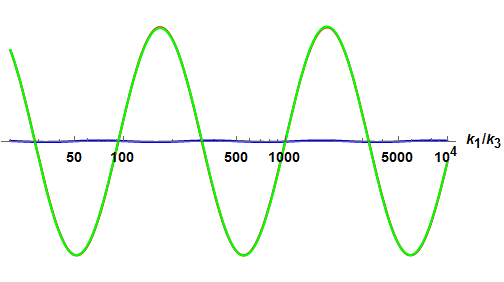}
  \end{center}
 \end{minipage}
 
  \caption{{\it{Left}} : Momentum ($k_1/k_3$) dependence of $(k_1/k_3)^2S $. The red, blue, and green lines denote $S_{11}$, $S_{22}$, and $S=S_{11}+S_{22}$, respectively. The masses are changed as $(m_1,m_2)/H=(2,2.01)$, $(2,2.1)$, and $(2,2.5)$ from top to bottom. In all of the figures, we take $g_{II}=\tilde{g}_{II}=1$ for $I=1,2$. {\it{Right}} : Detailed view of superposition. The constant offsets such as the last term in Eq.~$\eqref{case1S_11}$ are extracted to compare the three lines. In the last figure with $m_2/H=2.5$, the red and the green lines are overlapped.}
  \label{fig:one}
\end{figure}

In Figure~\ref{fig:one}, we show the behavior of the bispectrum~$(k_1/k_3)^2S$ as the function of $k_1/k_3$, by taking a universal value for the couplings, $g_{11}=g_{22}=\tilde{g}_{11}=\tilde{g}_{22}=1$. The red line shows the result of $S_{11}$, and the blue one denotes that of $S_{22}$. They correspond to the single isocurvaton cases with different masses. Their superposition is given by the green line.

We find that when the two masses are nearly degenerate, $S_{11}$ and $S_{22}$ take the almost same waveforms, and the superposition is just twice of $S_{11}$ (or $S_{22}$). See the upper right figure in Fig.~\ref{fig:one}. As the mass difference becomes larger, the amplitude of the heavier field ($S_{22}$) becomes smaller by the Boltzmann suppression~$\eqref{C_pm_asymp}$, and in this case, we have effectively obtained a single field signal (see the remaining two figures on the right). Therefore, we do not expect any signal with a waveform specific to two fields in this case.

\begin{figure}[t]
 \begin{minipage}{0.5\hsize}
  \begin{center}
   \includegraphics[width=70mm]{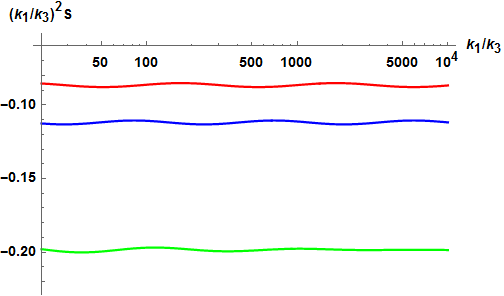}
  \end{center}
 \end{minipage}
 \begin{minipage}{0.5\hsize}
  \begin{center}
   \includegraphics[width=70mm]{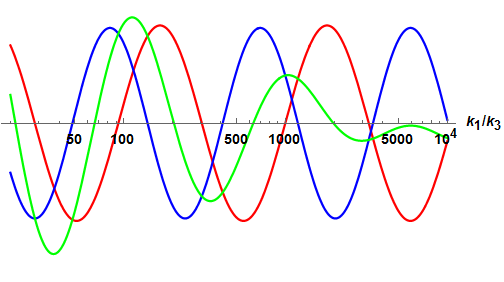}
  \end{center}
 \end{minipage}
 
  \begin{minipage}{0.5\hsize}
  \begin{center}
   \includegraphics[width=70mm]{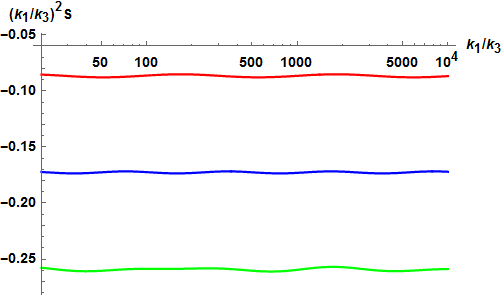}
  \end{center}
 \end{minipage}
  \begin{minipage}{0.5\hsize}
  \begin{center}
   \includegraphics[width=70mm]{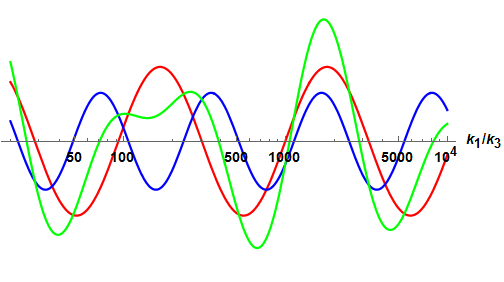}
  \end{center}
 \end{minipage}
 
\caption{The same figure with Fig.~\ref{fig:one} in the case with non-universal coupling constants. The masses and the couplings are set as $(m_1,m_2)/H=(2,2.1)$ and $(g_{11}\tilde{g}_{11},g_{22}\tilde{g}_{22})=(1,3)$ in the upper panel. Those of the lower one are $(m_1,m_2)/H=(2,2.5)$ and $(g_{11}\tilde{g}_{11},g_{22}\tilde{g}_{22})=(1,100)$, respectively.}
  \label{fig:two}
\end{figure}

Several non-trivial superpositions can be obtained for hierarchical choice of $g_{IJ}$ and $\tilde{g}_{IJ}$. Figure~\ref{fig:two} shows such non-trivial examples. 
One finds that even if the mass difference exists, we can still have a synergy effect, as long as the ratio $g_{22}\tilde{g}_{22}/g_{11}\tilde{g}_{11}$ is large so that the magnitude of $S_{22}$ is compatible to that of $S_{11}$. Compared to the previous case shown in Fig.~\ref{fig:one}, we can have a various type of superpositions since the frequency of each signal is much different for $\mu^1\neq \mu^2$. However, as the mass differences become larger, an extremely large hierarchy in $g_{IJ}$ or $\tilde{g}_{IJ}$ is required to realize such a non-trivial waveform by superposition. For example, for the case with $m_1=2H$ and $m_2=(3H,4H,5H,6H)$, we need to require $g_{22}\tilde{g}_{22}/g_{11}\tilde{g}_{11}\sim (10^4,10^8,10^{11},10^{14})$, respectively, in order for the contributions of $S_{11}$ and $S_{22}$ to be compatible.

\subsection{Effect of mixing}\label{Effect of mixing}
In this subsection, we discuss the effects of the mixing term~$S_{12}$. First, we discuss the case with $g_{IJ}=\tilde{g}_{IJ}=1$ for $I,J=1,2$. In the same way as the previous subsection, when $\mu^2\gg \mu^1$, the signals follow a relation $|S_{11}|\gg |S_{12}|\gg |S_{22}|$ by Eq.~$\eqref{C_pm_asymp}$. Thus,  
the total signal effectively becomes a single field one, and hence nothing interesting happens. Then, we focus on the degenerate case $\mu^1\sim \mu^2$ in the following, which is the most interesting case we deal with in this paper.

\begin{figure}[H]
 \begin{minipage}{0.5\hsize}
  \begin{center}
   \includegraphics[width=70mm]{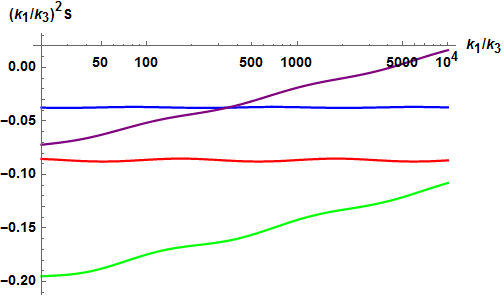}
  \end{center}
 \end{minipage}
 \begin{minipage}{0.5\hsize}
  \begin{center}
   \includegraphics[width=70mm]{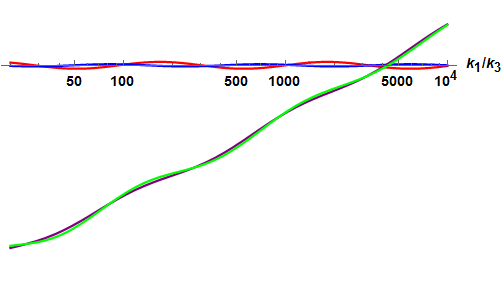}
  \end{center}
 \end{minipage}
  \begin{minipage}{0.5\hsize}
  \begin{center}
   \includegraphics[width=70mm]{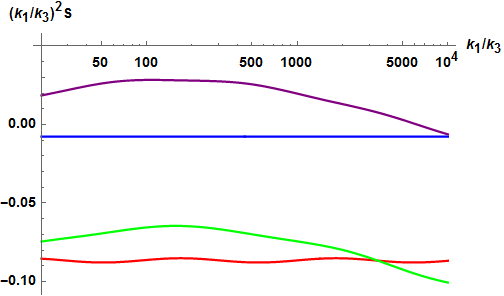}
  \end{center}
 \end{minipage}
 \begin{minipage}{0.5\hsize}
  \begin{center}
   \includegraphics[width=70mm]{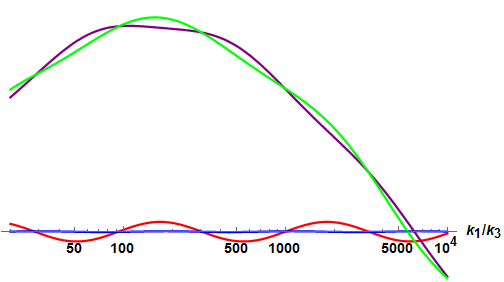}
  \end{center}
 \end{minipage}
  \begin{minipage}{0.5\hsize}
  \begin{center}
   \includegraphics[width=70mm]{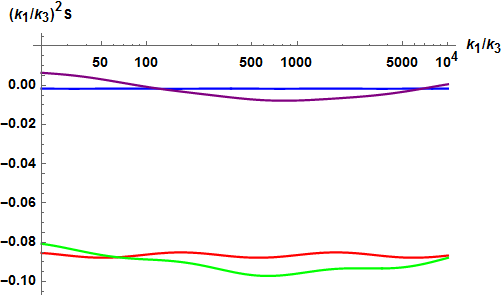}
  \end{center}
 \end{minipage}
 \begin{minipage}{0.5\hsize}
  \begin{center}
   \includegraphics[width=70mm]{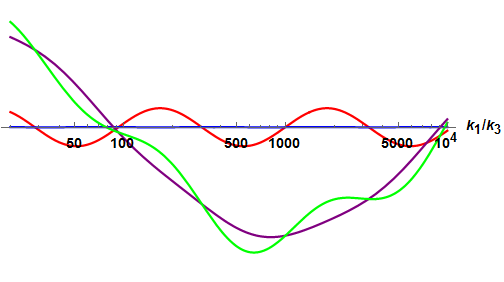}
  \end{center}
 \end{minipage}
 
\caption{The same figure with Fig.~\ref{fig:one} but the mixing term~$S_{12}$ included (purple line). We set $(m_1,m_2)/H=(2,2.1)$, $(m_1,m_2)/H=(2,2.3)$, and $(m_1,m_2)/H=(2,2.5)$ from top to bottom. The couplings are taken universally, $g_{IJ}=\tilde{g}_{IJ}=1$ for $I,J=1,2$. The right figures show that the waveforms (momentum dependence) of the total signal are mainly determined by the mixing term~$S_{12}$.   }
  \label{fig:three}
\end{figure}

When the two masses are nearly degenerate, $\mu^1=\mu$ and $\mu^2=\mu+\Delta$ with $0\leq \Delta\ll 1$, we find that the total waveform ($k$-dependence) of $S$ is dominated by $S_{12}$, while the magnitude of each signal including $k$-independent contribution satisfies $|S_{11}|\geq|S_{12}|\geq |S_{22}|$. One can confirm the statement above form Fig.~\ref{fig:three}, where we include the mixing term (purple line). The three kinds of the figures show the cases of $m_2/H=2.1, 2.3$, and $2.5$ with $m_1/H=2$ fixed, which corresponds to $\Delta=0.15, 0.42$, and $0.68$, respectively.
Obviously, the situation is much different from the case without a mixing term discussed in Sec.~\ref{Without mixing}, where we found that the total signal becomes a trivial superposition or almost the same form as the lighter one ($S_{11}$).

To analytically understand the reason why $S_{12}$ governs the $k$-dependence, remember the fact that $|C(i\mu^I,-i\mu^J)|\gg |C(i\mu^I,i\mu^J)|$ in the degenerate limit, as shown in Eq.~$\eqref{c+vsc-}$. Therefore, we can approximate the signal as
\begin{align}
\left(k_1/k_3\right)^2S_{11}/\mathcal{A}\sim &C(i\mu^1,-i \mu^1),\\
\left(k_1/k_3\right)^2 S_{12}/2\mathcal{A}\sim  &{\rm{Re}}C(i\mu^1,-i\mu^2){\rm{cos}}\left(\Delta \log\left(\frac{k_3}{2k_1} \right)\right)
-{\rm{Im}}C(i\mu^1,-i\mu^2){\rm{sin}}\left(\Delta\log\left(\frac{k_3}{2k_1} \right)\right),
\end{align}
where $\mathcal{A}\equiv -\frac{c_4}{4P_{\zeta}^{1/2}\pi^5}\frac{c_3}{H}$ is a universal prefactor. $S_{22}$ is given by $S_{11}$ with a replacement $\mu^1\rightarrow \mu^2$.
We find that $S_{11}$ and $S_{22}$ are constants in this limit, and therefore, the momentum dependence is governed by $S_{12}$. This observation is consistent with Fig.~\ref{fig:three}.\footnote{In the figure, we set $m_1=2H$ which corresponds to $\mu =1.3$. Thus, the condition $\mu\gg 1$ we used to derive Eq.~$\eqref{c+vsc-}$ is not satisfied, strictly speaking. However, the Stirling approximation~$\eqref{Stirling}$ is still valid up to $\mu \sim 1$, and the discussion here can be applied.} As can be seen from the expression, the signal is characterized by a large wavelength $1/\Delta$ with $\Delta\equiv \mu^2-\mu^1$, which is specific to the two isocurvaton model. Such a waveform with a relatively longer wavelength can be easily identified and useful for disentangling almost degenerate mass spectra, which is in sharp contrast with the case of collider experiments on earth, where the very precise energy resolution corresponding to the mass difference is required to disentangle almost degenerate mass spectra.
Note also that the two masses $m_1$ and $m_2$ do not have to be light to obtain this kind of signal. This is different from the single field case, where the signal with a large wavelength appears only when the isocurvaton mass is relatively light, $\mu \sim 0$ (or $m\sim 3H/2$).

In order to distinguish two different signals which come from multiple isocurvatons with degenerate masses and a single isocurvaton with $\mu\sim 0$, there are two points to be noticed. First, the coefficient $C$ determining the oscillation amplitude is different for each signal. The former one is characterized by Eq.~$\eqref{Cminus_asymp}$, while the latter one scales as $|C(i\mu,i\mu)|\sim 1/\mu^2$ for $\mu\sim 0$. Therefore, the aspect ratio of the waveform is much different each other. Second, in a multiple field case, there are small modulations on the large waveform, as can be seen from Fig.~\ref{fig:three}. This arises from the effect of superposition with $S_{11}$. Therefore, they are potentially distinguishable while it requires detailed analysis on the signals.

In summary, we find that when there exists the mixing term~$S_{12}$, the multiple isocurvatons with nearly degenerate masses produce a characteristic waveform on the bispectrum, whose wavelength is given by an inverse of $\Delta$ (mass difference, roughly). 

Before closing this section, let us briefly comment on the case where there exists a hierarchy in $g_{IJ}$ or $\tilde{g}_{IJ}$. As mentioned in the previous subsection, we can have a possibility for a non-trivial superposition (the signal of multiple fields in other words) even when $m_2>m_1$, but it requires a large hierarchy in the couplings, $g_{11}\tilde{g}_{11}$ and $g_{22}\tilde{g}_{22}$. When there are mixing terms, however, the situation becomes changed. Note that from Eq.~$\eqref{C_pm_asymp}$, $S_{12}$ is less suppressed than $S_{22}$. Therefore, the required hierarchy in $g_{IJ}$ or $\tilde{g}_{IJ}$ for $S_{11}$ and $S_{12}$ to be compatible is milder than that for $S_{11}$ and $S_{22}$ to be. In table~\ref{table_1}, taking $m_1=2H$, we show the required hierarchy for both cases with and without mixing terms. One can find that the relaxation occurs for the case with mixing.

\begin{table}[H]
\centering
\begin{tabular}{llll}
 &without mixing & &with mixing \\
$m_2=3H:$&$g_{22}\tilde{g}_{22}/g_{11}\tilde{g}_{11}\sim O(10^4)$&$\rightarrow$&$g_{12}\tilde{g}_{12}/g_{11}\tilde{g}_{11}\sim O(1)$\\
$m_2=4H:$&$g_{22}\tilde{g}_{22}/g_{11}\tilde{g}_{11}\sim O(10^8)$&$\rightarrow$&$g_{12}\tilde{g}_{12}/g_{11}\tilde{g}_{11}\sim O(10^2)$\\
$m_2=5H:$&$g_{22}\tilde{g}_{22}/g_{11}\tilde{g}_{11}\sim O(10^{11})$&$\rightarrow$&$g_{12}\tilde{g}_{12}/g_{11}\tilde{g}_{11}\sim O(10^4)$\\
$m_2=6H:$&$g_{22}\tilde{g}_{22}/g_{11}\tilde{g}_{11}\sim O(10^{14})$&$\rightarrow$&$g_{12}\tilde{g}_{12}/g_{11}\tilde{g}_{11}\sim O(10^6)$\\
\end{tabular}
\caption{hierarchy in $g_{IJ}$ and $\tilde{g}_{IJ}$ for non-trivial superposition.}
\label{table_1}
\end{table}

\section{Summary and discussion}\label{summary}
In this paper, we have investigated the effects of multiple scalar isocurvatons on the bispectrum of the primordial curvature perturbations, especially for the purpose of answering the question of how to disentangle the mass spectra of such isocurvatons. The contribution coming from the diagram in Fig.~\ref{fig_diagram} in the squeezed limit is explicitly evaluated. Taking the two fields case as an example, we classified the situation carefully in terms of presence or absence of mixing effects of two fields, magnitude relation of two masses, and hierarchy in the couplings, $g_{IJ}$ and $\tilde{g}_{IJ}$. As a main result, we found the following specific behavior of the bispectrum, which would be an evidence of multiple isocurvatons in the early universe.
\begin{itemize}
\item {\bf{Superposition}} : By superposition of two waves, several non-trivial waveforms which cannot be obtained from the single field case, can appear when there exists a hierarchy in $g_{IJ}$ or $\tilde{g}_{IJ}$. We further showed that the required hierarchy for such non-trivial superposition is different for the two cases, with or without the mixing term.
\item {\bf{Mixing effect}} : We observe a waveform with a large wavelength characterized by $|\Delta|^{-1}$ with $\Delta\equiv \sqrt{(m_1/H)^2-(3/2)^2}-\sqrt{(m_2/H)^2-(3/2)^2}$, if the mixing term between two isocurvatons exists and the two masses are nearly degenerate, $m_1\sim m_2$. 

\end{itemize}
Especially the second result is quite suggestive and useful. As mentioned in Introduction, in supersymmetric inflation scenario, we frequently encounter a situation where multiple scalar fields acquire nearly degenerate masses as large as just above the Hubble scale by supergravity corrections. Therefore we can expect this kind of signal appears in general and hence it is important to discuss how to resolve such degenerate mass spectra from the observations. Our results suggest the presence of a waveform with a relatively longer wavelength corresponding to the inverse of the mass difference roughly, which can be easily identified, while the very precise energy resolution corresponding to the mass difference is required to disentangle almost degenerate mass spectra in collider experiments on earth.

One of interesting applications is supergravity $\alpha$-attractor model~\cite{Kallosh:2013yoa}, which minimally contains three isocurvatons (a scalar partner of the inflaton and a complex stabilizer field) in addition to the inflaton. They have the Hubble induced masses with multiplication of some parameters including $\alpha$. Even if they are almost degenerate, we may easily disentangle the spectra by the mixing effect, and at the same time, obtain some restrictions on those parameters since they are degenerate. In this way, the mixing effect discussed in this work would be also useful to constrain UV models from observational signals.\footnote{In Ref.~\cite{Bartolo:2018hjc}, the single isocurvaton effect on the bispectrum in the supergravity $\alpha$-attractor model is discussed. }

Interestingly, the result could change when we consider other types of interactions between an inflaton and isocurvatons.
In Appendix~\ref{shift}, we perform the same analysis for the terms including derivative couplings, and find that the mixing effect seems not to appear in the other cases. Therefore, from these observations, it might be possible to distinguish whether the signal originates from a derivative coupling or a non-derivative coupling. However, it should also be kept in mind that our results, especially $D(i\mu^I,-i\mu^J) \sim 0$ in Eq.~\eqref{D_minus}, are based on the super-horizon (or late time) approximation. Although this approximation seems to be valid at $\mu^I\sim 1 (m^I\sim H)$~\cite{Chen:2016hrz} of our interest, to solidify the claim above requires a full numerical calculation of the bispectrum, as performed in Ref.~\cite{Chen:2015lza} for the case with a tree level diagram. If the result would be correct beyond the approximation, then the mixing effect could not only prove the existence of multi-particles around the Hubble scale and resolve their degenerate mass spectra, but also identify the form of interactions at the same time. Pursuing this would be an important research theme in the future.

Although multiple fields leave their traces on the bispectrum as the specific waveforms, the signal size is suppressed by the Boltzmann factor as usual, which makes the signal too small to be observed. In Refs.~\cite{Chen:2018xck,Wang:2019gbi,Wang:2020ioa,Bodas:2020yho}, however, it is pointed out that an introduction of effective chemical potentials via some kinds of dimension-$5$ derivative operators amplifies the signal, even when their masses are definitely larger than the Hubble scale. 
Thus, it would be interesting to apply our results to a system with chemical potentials. 

Another interesting direction is to extend our analysis to spinning particles. From the perspective of revealing the standard model background in the cosmological collider, it is important to understand properties of superposition of particles with different spins. Also, we expect that such extension would be useful for exploring a new physics such as supersymmetry. 
We leave all of them for future work.


\subsection*{Acknowledgements}
We would like to thank Xingang Chen, Yi Wang, and Zhong-Zhi Xianyu for helpful comments. M.\,Y. is supported in part by JSPS Grant-in-Aid for Scientific Research Number 18K18764 and JSPS Bilateral Open Partnership Joint Research Projects. 
\begin{appendix}

\section{Perturbation around inflationary background}\label{perturbation}
Here we specify the action describing the fluctuation around the inflationary background.

First of all, we decompose the four-dimensional metric by Arnowitt-Deser-Misner (ADM) decomposition~\cite{Arnowitt:1962hi}, 
\begin{align}
g_{\mu\nu}=\left(\begin{array}{cc}
-N^2+N^iN_i & N_{i} \\
N_j & h_{ij}
\end{array}\right) ,\ \ 
g^{\mu\nu}=\left(\begin{array}{cc}
-\frac{1}{N^2} & \frac{N^{i}}{N^2} \\
\frac{N^{j}}{N^2} & h^{ij}-\frac{N^{i}N^j}{N^2}
\end{array}\right),  
\end{align}
where $N$, $N^i$, and $h_{ij}$ are lapse, shift vector, and induced metric on constant-$t$ hypersurface, respectively. The three-dimensional index $i$ is raised/lowered by $h_{ij}$ and its inverse, $h^{ij}$.

Then, the fields and the metric are expanded as 
\begin{align}
&\phi=\phi_0(t)+\delta \phi, \ \ \sigma^I=0+\delta\sigma^I, \\
&N=1+\alpha, \ \ N_i=\partial_i\beta, \ \    h_{ij}=a^2\delta_{ij}, 
\end{align}
where we employed the so-called flat gauge (see Ref.~\cite{Wang:2013zva} for review), and omitted the vector- and tensor-perturbations. We use $\sigma^I$ instead of $\delta\sigma^I$ in the following.

\subsection{Free part}\label{perturbation_free}
The free part (quadratic order expansion) of Eq.~$\eqref{action}$ is given by
\begin{align}
\nonumber S^{(2)} = & \int d^4x a^3\biggl[ -2Ha^{-2}\alpha\partial^2\beta +(-3H^2+\frac{1}{2}\dot{\phi}_0^2)\alpha^2-a^{-2}\dot{\phi}_0\partial_i\beta\partial_i\delta\phi +\frac{1}{2}(\delta\dot{\phi})^2\\
\nonumber &-\dot{\phi}_0\alpha\delta\dot{\phi}-\frac{1}{2}a^{-2}(\partial\delta \phi)^2-\frac{1}{2}V_{\phi\phi}(\delta \phi)^2-V_{\phi}\alpha\delta\phi+\frac{1}{2}( \dot{\sigma}^I)^2-\frac{1}{2}a^{-2}(\partial  \sigma^I)^2\\
&-\frac{1}{2}U_{IJ}\sigma^I\sigma^J\biggr],\label{free}
\end{align}
where $\partial_i$ is a three-dimensional spatial derivative.

The E.O.Ms of $\alpha$ and $\beta$ give rise to the algebraic equations:
\begin{align}
&\alpha_{(1)}=\frac{\dot{\phi}_0}{2H}\delta \phi,\\
&\partial^2 \beta_{(1)}=-\frac{a^2}{2H}\left(6H^2\alpha_{(1)}+\dot{\phi}_0\delta \dot{\phi}-\dot{\phi}_0^2\alpha_{(1)}+V_{\phi}\delta\phi\right),
\end{align}
where the subscript $(1)$ explicitly denotes that they are the first order w.r.t. $\delta \phi$. By substituting them into Eq.~$\eqref{free}$, we obtain 
Eq.~$\eqref{free2}$.

\subsection{Interaction}\label{perturbation_int}

Here we focus on three and four point vertices which arise from Eq.~$\eqref{int_f}$ or \footnote{The other sectors than $S_{\rm{int}}$ also produce these vertices, but we focus on the terms originating from $S_{\rm{int}}$ in this paper.} 
\begin{align}
S_{\rm{int}}(\phi,\sigma)=-\int d^{4} x \sqrt{-g}f(\phi, X)  g_{IJ}\sigma^I\sigma^J. \label{int}
\end{align}
Now we expand Eq.~$\eqref{int}$ up to quartic order of perturbations and obtain,
\begin{align}
\nonumber \mathcal{L}_{\rm{int}}=&- g_{IJ}\sigma^I\sigma^Ja^3\biggl[ f_0+f_{\phi}\delta \phi -2f_X\dot{\phi}_0\delta \dot{\phi} +2f_X\dot{\phi}_0^2\alpha+f_0\alpha \\
\nonumber &+\frac{1}{2}f_{\phi\phi}(\delta\phi)^2+(-f_X+2f_{XX}\dot{\phi}_0^2)(\delta\dot{\phi})^2+(2f_X\dot{\phi}_0-2f_{XX}\dot{\phi}_0^3)\alpha\delta\dot{\phi}\\
\nonumber &+(-f_X\dot{\phi}_0^2+2f_{XX}\dot{\phi}_0^4)\alpha^2+2f_Xa^{-2}\dot{\phi}_0\partial_i\delta\phi\partial_i\beta +f_Xa^{-2}(\partial\delta\phi)^2\\
&-2f_{\phi X}\dot{\phi}_0\delta\phi\delta\dot{\phi}+(f_{\phi}+2f_{\phi X}\dot{\phi}_0^2)\alpha \delta\phi\biggr],\label{int_expanded}
\end{align}
where $f_0\equiv f(\phi_0, X_0)$ and the derivatives~$f_{\phi}, f_{X},\cdots$ are evaluated at the background.

Then, we eliminate auxiliary fields~$\alpha$ and~$\beta$. Note that when we discuss quartic order perturbations, we need to take into account the solution of $\alpha$ and $\beta$ up to the second order w.r.t. the perturbations, $\delta\phi$ and $\sigma^I$~\cite{Maldacena:2002vr}, which requires the cubic order expansion of the total action in addition to Eq.~$\eqref{int_expanded}$. The expanded action other than Eq.~$\eqref{int_expanded}$ is given by 
\begin{align}
\nonumber S^{(3)}|_{{\rm{without}}\ S_{\rm{int}}}=&\int d^4x a^3\biggl[\frac{1}{2}a^{-4}\alpha(\partial^2\beta)^2-\frac{1}{2}a^{-4}\alpha(\partial_i\partial_j\beta)^2+2Ha^{-2}\alpha^2\partial^2\beta\\
\nonumber &+3H^2\alpha^3-a^{-2}\delta\dot{\phi}\partial_i\beta\partial_i\delta\phi-a^{-2}\delta\dot{\sigma}^I\partial_i\beta\partial_i\delta\sigma^I+a^{-2}\dot{\phi}_0\alpha \partial_i\beta\partial_i\delta\phi\\
\nonumber &-\frac{1}{2}\alpha(\delta\dot{\phi})^2-\frac{1}{2}\alpha(\dot{\sigma}^I)^2+\dot{\phi}_0\alpha^2\delta\dot{\phi}-\frac{1}{2}(\dot{\phi}_0)^2\alpha^3-\frac{1}{2}a^{-2}\alpha(\partial\delta\phi)^2\\
\nonumber &-\frac{1}{2}a^{-2}\alpha(\partial\sigma^I)^2-\frac{1}{6}V_{\phi\phi\phi}(\delta\phi)^3-\frac{1}{2}V_{\phi\phi}\alpha(\delta\phi)^2-\frac{1}{6}U_{IJK}\sigma^I\sigma^J\sigma^K\\
&-\frac{1}{2}U_{IJ}\alpha\sigma^I\sigma^J\biggr].
\end{align}

In the following, we assume a slow-roll inflation scenario, and extract terms up to $O(\sqrt{\epsilon})$ where $\epsilon\equiv -\dot{H}/H^2=\dot{\phi}_0^2/2H^2\ll 1$ is a slow-roll parameter. Moreover, we focus on $\delta\phi$-$\sigma^2$ and $\delta\phi^2$-$\sigma^2$ type interactions.
Then, the relevant parts of the solutions for $\alpha=\alpha_{(1)}+\alpha_{(2)}+\cdots $ are given by
\begin{align}
&\alpha_{(1)}=\frac{\dot{\phi}_0}{2H}\delta \phi,\\
&\partial^2 \alpha_{(2)}=\frac{1}{2H}\partial_i\left(\partial_i\delta \phi \delta \dot{\phi}\right)+\cdots,
\end{align}
where $\alpha_{(n)}$ are the $n$-th order solutions.\footnote{We do not need the solution~$\beta_{(n)}$ under the slow-roll approximation up to $\sqrt{\epsilon}$.}  
The ellipsis in the solutions denotes the terms containing $\sigma^I$ and higher order of $\epsilon$ (see Refs.~\cite{Seery:2006vu,Arroja:2008ga} for the complete expression).

By inserting the solutions into Eq.~$\eqref{int_expanded}$, we find 
\begin{align}
\nonumber \mathcal{L}_{\rm{int}} =  &- g_{IJ}\sigma^I\sigma^Ja^3\biggl[f_0+\left(f_{\phi}+f_0\frac{\dot{\phi}_0}{2H}\right)\delta \phi -2f_X\dot{\phi}_0\delta\dot{\phi}\\
\nonumber &+\frac{1}{2}\left(f_{\phi\phi}+f_{\phi}\frac{\dot{\phi}_0}{H}\right)(\delta\phi)^2+f_Xg_{0}^{\mu\nu}\partial_{\mu}\delta\phi\partial_{\nu}\delta\phi+\frac{f_0}{2H}\partial^{-2}\partial_i(\partial_i\delta\phi \delta\dot{\phi}) \\
&-2f_{\phi X}\dot{\phi}_0\delta\phi\delta\dot{\phi}\biggr]+\cdots,\label{int_234}
\end{align}
where $g_0^{\mu\nu}={\rm{diag}}(-1,a^{-2},a^{-2},a^{-2})$, and the ellipsis denotes irrelevant and subdominant terms for our discussion. While the first term contributes to the free part~$\eqref{free}$, the discussion remains unchanged by replacing $U_{IJ}\rightarrow U_{IJ}+2g_{IJ}f_0$.

\section{Derivation of bispectrum}\label{derivation}
In this section, we derive Eq.~$\eqref{result}$ (see Ref.~\cite{Chen:2016hrz} for more detail). 

First, applying the Schwinger-Keldysh formalism~\cite{Chen:2017ryl} to the system with the interactions~$\eqref{L_3}$ and $\eqref{L_4}$, we obtain the 1-loop correction on the bispectrum from $\sigma^I$ as 
\begin{align}
\nonumber &\left\langle\delta \phi_{\bm{k}_{1}} \delta \phi_{\bm{k}_{2}} \delta \phi_{\bm{k}_3}\right\rangle '  \\
\nonumber =&-\frac{c_3c_4}{H^2}\frac{1}{4(k_1k_2k_3)^3}\sum_{I,J}g_{IJ}\tilde{g}_{IJ}\int^0_{-\infty}\frac{d\tau_1}{\tau_1^4}\int^0_{-\infty}\frac{d\tau_2}{\tau_2^4}\\
\nonumber &\times \sum_{a,b=\pm}(ab)e^{iak_3\tau_1+ib(k_1+k_2)\tau_2}(1-iak_3\tau_1)(1-ibk_2\tau_2)(1-ibk_1\tau_2)\\
&\times \int \frac{d^3{\bf{k}}}{(2\pi)^3}G^I_{ab}(k;\tau_1,\tau_2)G^J_{ab}(|{\bf{k}}_3+{\bf{k}}|;\tau_1,\tau_2)+{\rm{5\ permutations\ of\ }} {\bf{k}}_1,{\bf{k}}_2\ {\rm{and}}\ {\bf{k}}_3,\label{SK}
\end{align}
where the indices $(a,b)$ take either $+1$ or $-1$ according to the $+$ and $-$ contours in Schwinger-Keldysh formalism. The propagator $G^I_{ab}$ of $\sigma^I$ is explicitly given by
\begin{align}
&G^I_{-+}(k;\tau_1,\tau_2)=v^I_k(\tau_1)v_k^{I*}(\tau_2),\\
&G^I_{++}(k;\tau_1,\tau_2)=\theta (\tau_1-\tau_2)G^I_{-+}(k;\tau_1,\tau_2)+\theta (\tau_2-\tau_1)G^I_{+-}(k;\tau_1,\tau_2),
\end{align}
and, $G^I_{+-}(k;\tau_1,\tau_2)=\left(G^I_{-+}(k;\tau_1,\tau_2)\right)^*$ and $G^I_{--}(k;\tau_1,\tau_2)=\left(G^I_{++}(k;\tau_1,\tau_2)\right)^*$. $\theta(\tau)$ denotes a unit step function.

In the following, we focus on the two terms including ${\bf{k}}_3$ in the loop integral in Eq.~$\eqref{SK}$, which correspond to the diagram, Fig.~\ref{fig_diagram}, since they produce an oscillating behavior in the squeezed limit $(k_1\sim k_2\gg k_3)$. By using 
\begin{align}
G^I_{ab}(x_1,x_2)=\int \frac{d^3{\bf{k}}}{(2\pi)^3}G^I_{ab}(k;\tau_1,\tau_2)e^{-i{\bf{k}}\cdot{\bf{X}}},
\end{align}
with ${\bf{X}}\equiv {\bf{x}}_{1}-{\bf{x}}_{2}$, we can rewrite the loop integral as
\begin{align}
{\rm{Loop}}\left(\equiv \int \frac{d^3{\bf{k}}}{(2\pi)^3}G^I_{ab}(k;\tau_1,\tau_2)G^J_{ab}(|{\bf{k}}_3+{\bf{k}}|;\tau_1,\tau_2)\right)=\int d^3{\bf{X}}G^I_{ab}(x_1,x_2)G^J_{ab}(x_1,x_2)e^{-i{\bf{k}}_{3}\cdot{\bf{X}}}. 
\end{align}
It is difficult to perform the integration, but analytic expression can be obtained by approximating the propagators in the super-horizon limit,~$|k\tau_1|,|k\tau_2| \leq 1$, \cite{Chen:2016hrz}. 
Under this approximation, $G^I_{ab}$ produce the same result independently of the $a,b$ indices,
\begin{align}
G^I_{ab}(x_1,x_2)  \simeq \frac{H^2}{4\pi^{5/2}}\biggl[X^{-3+2\nu^I}(\tau_1\tau_2)^{\frac{3}{2}-\nu^I}\Gamma(\nu^I)\Gamma(\frac{3}{2}-\nu^I)+(\nu^I\rightarrow -\nu^I)\biggr],
\end{align}
and we obtain
\begin{align}
\nonumber {\rm{Loop}}\simeq
&\ \frac{H^4}{4\pi^4}\biggl[(k_3)^{3-2(\nu^I+\nu^J)}(\tau_1\tau_2)^{3-(\nu^I+\nu^J)}I(\nu^I,\nu^J)+(\nu^I\rightarrow \nu^I,\nu^J\rightarrow -\nu^J)\\
&+(\nu^I\rightarrow -\nu^I,\nu^J\rightarrow \nu^J)+(\nu^I\rightarrow -\nu^I,\nu^J\rightarrow -\nu^J)\biggr], \label{loop}
\end{align}
where
\begin{align}
 I(\nu^I,\nu^J)\equiv \Gamma(\nu^I)\Gamma(\nu^J)\Gamma(\frac{3}{2}-\nu^I)\Gamma(\frac{3}{2}+\nu^I)\Gamma(-4+2(\nu^I+\nu^J)) {\rm{sin}}(\pi (\nu^I+\nu^J)).   
\end{align}

Finally, by performing the time integration $\tau_1$ and $\tau_2$, we obtain the following approximated expression,
\begin{align}
\nonumber &\left\langle\delta \phi_{\bm{k}_{1}} \delta \phi_{\bm{k}_{2}} \delta \phi_{\bm{k}_3}\right\rangle '  \\
\nonumber \simeq &\frac{H^2c_3c_4}{2 \pi^4}\sum_{I,J}g_{IJ}\tilde{g}_{IJ}\frac{1}{(k_1 k_2)^3}\biggl\{ k_3^{-(\nu^I+\nu^J)}(k_1+k_2)^{\nu^I+\nu^J}\\
\nonumber &\times I(\nu^I,\nu^J)\Gamma^2(-\nu^I-\nu^J) {\rm{sin}}^2(\frac{\pi}{2}(\nu^I+\nu^J))(1-\nu^I-\nu^J)^2\left(1-\frac{k_1k_2}{(k_1+k_2)^2}(\nu^I+\nu^J)\right)\\
&+(\nu^I\rightarrow \nu^I,\nu^J\rightarrow -\nu^J)+(\nu^I\rightarrow -\nu^I,\nu^J\rightarrow \nu^J)+(\nu^I\rightarrow -\nu^I,\nu^J\rightarrow -\nu^J)\biggr\}, \label{pre_result}
\end{align}
which leads to Eq.~$\eqref{result}$ in the squeezed limit, $k_3\ll k_1\sim k_2$.

\section{Other interactions}\label{shift}

Here we discuss other type of vertices constructed from Eq.~$\eqref{int_234}$. Practically, a case where the inflaton respects the shift symmetry, $\phi\rightarrow \phi+c$ with $c \in \mathbb{R}$, is another important example. In this case, only the two terms proportional to $f_X$ in Eq.~$\eqref{int_234}$ survive, which are written here again:
\begin{align}
&\mathcal{L}_{\rm{int}}^3=2a^3f_X\dot{\phi}_0\delta\dot{\phi}g_{IJ}\sigma^I\sigma^J,\\
&\mathcal{L}_{\rm{int}}^4=-a^3f_X (\partial_{\mu}\delta\phi)^2g_{IJ}\sigma^I\sigma^J,
\end{align}
with $(\partial_{\mu}\delta\phi)^2=g_{0}^{\mu\nu}\partial_{\mu}\delta\phi\partial_{\nu}\delta\phi$. In the same way as the main text, we parametrize them as 
\begin{align}
&\mathcal{L}_{\rm{int}}^3=a^3d_3\delta\dot{\phi}g_{IJ}\sigma^I\sigma^J,\label{int_3_shift}\\
&\mathcal{L}_{\rm{int}}^4=a^3d_4(\partial_{\mu}\delta\phi)^2\tilde{g}_{IJ}\sigma^I\sigma^J,\label{int_4_shift}
\end{align}
where $d_3$ and $d_4$ are two independent parameters, and $\tilde{g}_{IJ}$ can be different from $g_{IJ}$ in general. Note that $d_3$ and $d_4H^2$ are dimensionless. 

The interactions~$\eqref{int_3_shift}$ and $\eqref{int_4_shift}$ produce a similar diagram to Fig.~\ref{fig_diagram} but the vertices are replaced accordingly. We can evaluate the relevant parts of the bispectrum as 
\begin{align}
\nonumber &\left\langle\delta \phi_{\bm{k}_{1}} \delta \phi_{\bm{k}_{2}} \delta \phi_{\bm{k}_3}\right\rangle '  \\
\nonumber =&-Hd_3d_4\frac{k_3^2}{2(k_1k_2k_3)^3}\sum_{I,J}g_{IJ}\tilde{g}_{IJ}\int^0_{-\infty}\frac{d\tau_1}{\tau_1^2}\int^0_{-\infty}\frac{d\tau_2}{\tau_2^2}\\
\nonumber &\times \sum_{a,b=\pm}(ab)e^{iak_3\tau_1+ib(k_1+k_2)\tau_2}\left(k_1^2k_2^2\tau_2^2+{\bf{k}}_1\cdot {\bf{k}}_2(1-ibk_1\tau_2)(1-ibk_2\tau_2)\right)\\
&\times \int \frac{d^3{\bf{k}}}{(2\pi)^3}G^I_{ab}(k;\tau_1,\tau_2)G^J_{ab}(|{\bf{k}}_3+{\bf{k}}|;\tau_1,\tau_2).
\end{align}
After repeating the same procedure as in Appendix~\ref{derivation}, we obtain
\begin{align}
\nonumber &\left\langle\delta \phi_{\bm{k}_{1}} \delta \phi_{\bm{k}_{2}} \delta \phi_{\bm{k}_3}\right\rangle '  \\
\nonumber =&\frac{H^5d_3d_4}{4 \pi^4}\sum_{I,J}g_{IJ}\tilde{g}_{IJ}\frac{1}{k_1^6}\biggl\{ D(\nu^I,\nu^J)\left(\frac{k_3}{2k_1}\right)^{-(\nu^I+\nu^J)}+(\nu^I\rightarrow \nu^I,\nu^J\rightarrow -\nu^J)\\
&+(\nu^I\rightarrow -\nu^I,\nu^J\rightarrow \nu^J)+(\nu^I\rightarrow -\nu^I,\nu^J\rightarrow -\nu^J)\biggr\}, \label{result_shift}
\end{align}
where
\begin{align}
\nonumber D(\nu^I,\nu^J)=&(3-\nu^I-\nu^J)\Gamma^2(2-\nu^I-\nu^J)\Gamma(\nu^I)\Gamma(\nu^J)\\
\nonumber &\times \Gamma(\frac{3}{2}-\nu^I)\Gamma(\frac{3}{2}-\nu^J)\Gamma(-4+2(\nu^I+\nu^J))\\
&\times {\rm{cos}}(\frac{\pi}{2}(\nu^I+\nu^J)){\rm{sin}}^3(\frac{\pi}{2}(\nu^I+\nu^J)).\label{D}
\end{align}
The shape function in this case is given by
\begin{align}
 \nonumber S=&-\frac{d_3}{2P_{\zeta}^{1/2}\pi^5}(d_4H^2)\sum_{I,J}g_{IJ}\tilde{g}_{IJ}\biggl\{ D(\nu^I,\nu^J)\left(\frac{k_3}{2k_1}\right)^{2-(\nu^I+\nu^J)}+(\nu^I\rightarrow \nu^I,\nu^J\rightarrow -\nu^J)\\
&+(\nu^I\rightarrow -\nu^I,\nu^J\rightarrow \nu^J)+(\nu^I\rightarrow -\nu^I,\nu^J\rightarrow -\nu^J)\biggr\}.
\label{shape_shift}
\end{align}

Let us see whether the mixing effect occurs or not in this situation. 
 In the degenerate limit~$\nu^I\equiv i \mu^I\sim \nu^J\equiv i\mu^J$, we have
\begin{align}
D(i\mu^I,-i\mu^J) \sim 0,  \label{D_minus} 
\end{align}
while 
\begin{align}
|D(i\mu^I,i\mu^J)| \sim \frac{\pi^{7/2}}{2^5\sqrt{2}}\mu^{1/2}e^{-2\pi\mu},   
\end{align}
where we have set $\mu^I\sim \mu^J=\mu \gg 1$. Obviously, these features are in contrast to those of Eqs.~$\eqref{Cplus_asymp}$ and~$\eqref{Cminus_asymp}$ in the non-derivative coupling case, where $|C(i\mu^I,-i\mu^J)|\gg |C(i\mu^I,i\mu^J)|$ holds in this limit. Remember that the mixing effect appears because $C(i\mu^I,-i\mu^J)$ dominates over the $k$-dependent part of the signal. On the other hand, $D(i\mu^I,-i\mu^J)$ never dominates the waveform in the degenerate limit, and therefore, we conclude that such behavior does not arise in this case.

This result can be applied to the other remaining cases, which can be made from Eq.~$\eqref{int_234}$. There, we have $2$ ($4$) types of three (four) point vertices, which totally gives $8$ combinations as $\mathcal{L}_{\rm{int}}^3+\mathcal{L}_{\rm{int}}^4$. We found that, in any other combination except for Eqs.~$\eqref{L_3}$ and $\eqref{L_4}$ (the combination discussed in the main text), the terms corresponding to $D(i\mu^I,-i\mu^J)$ behave as Eq.~$\eqref{D_minus}$ in the degenerate mass limit, and thus there is no mixing effect. Although we do not perform the detailed analysis for the other cases, the statement above can be technically understood as follows.
First, note that the loop contribution is common for all cases, given in Eq.~$\eqref{loop}$, and the  difference essentially comes from the number of time derivative on the inflaton fluctuation $\delta \phi$. Thus, it is reasonable to parametrize the interactions as 
\begin{align}
&\mathcal{L}_{\rm{int}}^3= a^3\delta \phi^{(n)} g_{IJ}\sigma^I\sigma^J,\\
&\mathcal{L}_{\rm{int}}^4= a^3\delta \phi^{(m)}\delta \phi^{(\ell)} \tilde{g}_{IJ}\sigma^I\sigma^J,
\end{align}
where $(n)$ on $\delta\phi$ denotes its $n$-th time derivative. 
By extracting the parts with $\tau_{1,2}$ dependence in the bispectrum, its rough structure is summarized as
\begin{align}
\nonumber &\int_{-\infty}^0d\tau_1\int_{-\infty}^0d\tau_2\sum_{a,b=\pm}(ab)\tau_1^{-4+n}\tau_2^{-4+m+\ell}(\tau_1\tau_2)^{3-\nu^I-\nu^J}\frac{\partial^n}{\partial \tau_1^n}u^{(a)}_{k_3} (\tau_1) \frac{\partial^m}{\partial \tau_2^m}u^{(b)}_{k_1}(\tau_2)\frac{\partial^{\ell}}{\partial \tau_2^{\ell}}u^{(b)}_{k_2}(\tau_2)\\
&+(\nu^I\rightarrow \nu^I,\nu^J\rightarrow -\nu^J)+(\nu^I\rightarrow -\nu^I,\nu^J\rightarrow \nu^J)+(\nu^I\rightarrow -\nu^I,\nu^J\rightarrow -\nu^J),
\label{t-dependence}
\end{align}
where $u^{(a)}_{k}(\tau)\equiv (1-iak\tau)e^{iak\tau}$, i.e., $u^{(-)}=u$ and $u^{(+)}=u^*$. The factors $-4+n$ and $-4+m+\ell$ in the power of $\tau_{1,2}$ come from the measure $\sqrt{-g}=a^3$ and the coordinate transformation $dt=ad\tau$ with $a=-1/H\tau$. The factor $3-\nu^I-\nu^J$ is originated from the loop integral (see Eq.~$\eqref{loop}$). Finally, the derivative of the mode function is given explicitly by  
\begin{align}
\frac{\partial^n}{\partial \tau^n}u^{(a)}_{k}(\tau)=-e^{i\frac{\pi}{2}n} (ak)^n(n-1+iak\tau)e^{iak\tau}.    
\end{align}
Inserting this expression into Eq.~$\eqref{t-dependence}$, and integrating it over $\tau_1$ and $\tau_2$, we find that the second and the third terms in Eq.~$\eqref{t-dependence}$ are proportional to
\begin{align}
\Gamma(n\mp \nu^I\pm \nu^J)\Gamma(m+\ell\mp\nu^I\pm \nu^J){\rm{sin}}^2\frac{\pi}{2}(\pm\nu^I\mp\nu^J),
\end{align}
which contributes to $C(\pm\nu^I,\mp\nu^J)$ or $D(\pm\nu^I,\mp\nu^J)$ (the double sign applies in the same order as written). Then, one can find that these factors vanish in the degenerate mass limit~$\nu^I\sim \nu^J$ unless $n=m=\ell=0$, which corresponds to the non-derivative vertices discussed in the main text. Therefore, we conclude that the mixing effect (signal with a large wavelength in the degenerate mass limit) is specific to the non-derivative interactions under the super-horizon approximation.

\section{Supplementary discussion}\label{Origin}
By changing the field basis, here we discuss 
\begin{itemize}
  \item (\ref{H}) how the hierarchical structure in $g_{IJ}$ for the non-trivial superposition could be originated,
  \item (\ref{I}) how the mixing effect can be interpreted in an interaction-diagonalized basis.
\end{itemize}

\subsection{Hierarchical structure in $g_{IJ}$ from mass mixing}\label{H}
In the main text, we find that a non-trivial superposition can appear when there exists a hierarchical structure in the coupling constant~$g_{IJ}$ (see Fig.~\ref{fig:two}). There, we took $U_{IJ}$ as diagonal from the beginning, and assumed some hierarchies in $g_{IJ}$. Here we show that such hierarchy can be effectively realized from a mass mixing of the isocurvatons in $U_{IJ}$. Again, we focus on the case with $n=2$.

The mass matrix~$U_{IJ}$ includes an off diagonal element~$U_{12}$ in general. Then, we can diagonalize $U_{IJ}$ as 
\begin{align}
U_{IJ}\sigma^I\sigma^J =\sum_{I=1,2}M_I^2(\tilde{\sigma}^{I})^2,   
\end{align}
where
\begin{align}
M_{1,2}^2\equiv \frac{1}{2}\left(U_{11}+U_{22}\pm \sqrt{(U_{11}-U_{22})^2+4U_{12}^2}\right),
\end{align}
and
\begin{align}
\left(\begin{array}{l}
\tilde{\sigma}^{1} \\
\tilde{\sigma}^{2}
\end{array}\right)=\left(\begin{array}{ll}
\text {cos} \theta & \text {sin} \theta \\
-\text {sin} \theta & \text {cos}\theta
\end{array}\right)\left(\begin{array}{l}
\sigma^{1} \\
\sigma^{2}
\end{array}\right)\ \ {\rm{with}}\ \  {\rm{tan}}2\theta =\frac{2U_{12}}{U_{11}-U_{22}}.
\end{align}

In this new basis, one can rewrite the interaction as
\begin{align}
fg_{IJ}\sigma^I\sigma^J=f\tilde{g}_{IJ}\tilde{\sigma}^I\tilde{\sigma}^J,
\end{align}
where 
\begin{align}
\tilde{g}_{IJ}=\left(\begin{array}{cc}
g_{11} \cos^{2} \theta+g_{22} \sin ^{2} \theta+g_{12} \sin 2 \theta &\frac{1}{2} (g_{22}-g_{11}) \sin 2\theta+g_{12} \cos 2 \theta \\
\ast& g_{11} \sin ^{2} \theta+g_{22} \cos ^{2} \theta-g_{12} \sin 2 \theta
\end{array}\right).
\end{align}
In particular, when $g_{IJ}= g{\bf{1}}_{2\times 2}$ for $I=1,2$, we obtain
\begin{align}
\tilde{g}_{IJ}=g\left(\begin{array}{cc}
1+\frac{2 U_{12}}{\sqrt{\left(U_{11}-U_{22}\right)^{2}+4 U_{12}^{2}}} & \frac{U_{11}-U_{22}}{\sqrt{\left(U_{11}-U_{22}\right)^{2}+4 U_{12}^{2}}} \\
\ast& 1-\frac{2 U_{12}}{\sqrt{\left(U_{11}-U_{22}\right)^{2}+4 U_{12}^{2}}}
\end{array}\right).
\end{align}
Therefore, even if the coupling constants $g_{IJ}$ are the same order in the original basis, we could obtain a large hierarchy between the effective coupling constants $\tilde{g}_{11}$ and $\tilde{g}_{22}$, when $U_{11}\sim U_{22}$ and $U_{12}\neq 0$.

\subsection{Interaction diagonalized basis}\label{I}
In section~\ref{Effect of mixing}, we saw that the total waveform is governed by some terms in the mixing $S_{12}$ with a wavelength $1/|\mu^2-\mu^1|$ (the inverse of the mass difference roughly), when the masses of the two isocurvatons are nearly degenerate, which play an important role to disentangle the spectra. Of course, this feature is specific to the multi isocurvatons model since the notion of the mass difference only exists in the case with $n\geq 2$. In this subsection, we show that the specific mode characterized by the mass difference cannot be mimicked by a single isocurvaton exchange from a different point of view. To do so, let us diagonalize the interactions $fg_{IJ}\sigma^I\sigma^J$ with respect to $\sigma^I$ this time, so that there is no mixing in the coupling constant in the new basis. Then, we obtain
\begin{align}
fg_{IJ}\sigma^I\sigma^J=\sum_{I=1,2}f\hat{g}_I(\hat{\sigma}^I)^2,\label{hat_g}
\end{align}
where 
\begin{align}
\hat{g}_I\equiv \frac{1}{2}\left(g_{11}+g_{22}\pm \sqrt{(g_{11}-g_{22})^2+4g_{12}^2}\right),
\end{align}
and
\begin{align}
\left(\begin{array}{l}
\hat{\sigma}^{1} \\
\hat{\sigma}^{2}
\end{array}\right)=\left(\begin{array}{ll}
\text{cos} \varphi& \text{sin} \varphi \\
-\text{sin} \varphi & \text{cos} \varphi
\end{array}\right)\left(\begin{array}{l}
\sigma^{1} \\
\sigma^{2}
\end{array}\right)\ \ {\rm{with}}\ \  {\rm{tan}}2\varphi =\frac{2g_{12}}{g_{11}-g_{22}}.
\end{align}
Then, the original (diagonal) mass matrix~$U_{IJ}$ are changed as
\begin{align}
U_{IJ}\sigma^I\sigma^J=\hat{U}_{IJ}\hat{\sigma}^I\hat{\sigma}^J,
\end{align}
where
\begin{align}
\hat{U}_{IJ}=\left(\begin{array}{cc}
U_{11}{\rm{cos}}^2\varphi+U_{22}{\rm{sin}}^2\varphi & \frac{1}{2}(U_{22}-U_{11}) {\rm{sin}}2\varphi\\
\ast&U_{11}{\rm{sin}}^2\varphi+U_{22}{\rm{cos}}^2\varphi 
\end{array}\right). \label{hat_mass}
\end{align}
Note that two isocurvatons $\hat{\sigma}^I$ individually couple to inflaton by Eq.~$\eqref{hat_g}$ without a mixing term, but they rather have a mass mixing as shown in Eq.~$\eqref{hat_mass}$. Since ${\rm{sin}}2\varphi \propto g_{12}$, one can see that the mixing of $g_{IJ}$ is now converted to that of $\hat{U}_{IJ}$.
 
From this expression, we can understand that the oscillating terms with the frequency~$|\mu^2-\mu^1|$ in $S_{12}$ (the third and fourth terms in Eq.~$\eqref{case1S_12}$) are related to the mass mixing $\hat{U}_{12}\propto U_{22}-U_{11}$ in the new basis. Therefore, we can conclude that the long waveform in $S_{12}$ or the mixing effect arising from it can be interpreted in the new basis as a consequence of the transfer term of $\sigma^1\leftrightarrow\sigma^2$ with a coupling constant $U_{22}-U_{11}$. Again, this kind of conversion is specific to the multi isocurvatons model and cannot be mimicked by a single isocurvaton exchange.\footnote{Note that one can proceed to calculate the bispectrum in the new basis, for example by treating the mass mixing $\hat{U}_{12}$ as a small perturbation. In this sense, our treatment in the main text allows us to take into account such effect non-perturbatively.}

\end{appendix}



\begin{thebibliography}{99}


\bibitem{Chen:2009we}
X.~Chen and Y.~Wang,
``Large non-Gaussianities with Intermediate Shapes from Quasi-Single Field Inflation,''
Phys. Rev. D \textbf{81}, 063511 (2010)
[arXiv:0909.0496 [astro-ph.CO]].

\bibitem{Chen:2009zp}
X.~Chen and Y.~Wang,
``Quasi-Single Field Inflation and Non-Gaussianities,''
JCAP \textbf{04}, 027 (2010)
[arXiv:0911.3380 [hep-th]].

\bibitem{Baumann:2011nk}
D.~Baumann and D.~Green,
``Signatures of Supersymmetry from the Early Universe,''
Phys. Rev. D \textbf{85}, 103520 (2012)
[arXiv:1109.0292 [hep-th]].

\bibitem{Assassi:2012zq}
V.~Assassi, D.~Baumann and D.~Green,
``On Soft Limits of Inflationary Correlation Functions,''
JCAP \textbf{11}, 047 (2012)
[arXiv:1204.4207 [hep-th]].

\bibitem{Sefusatti:2012ye}
E.~Sefusatti, J.~R.~Fergusson, X.~Chen and E.~P.~S.~Shellard,
``Effects and Detectability of Quasi-Single Field Inflation in the Large-Scale Structure and Cosmic Microwave Background,''
JCAP \textbf{08}, 033 (2012)
[arXiv:1204.6318 [astro-ph.CO]].

\bibitem{Norena:2012yi}
J.~Norena, L.~Verde, G.~Barenboim and C.~Bosch,
``Prospects for constraining the shape of non-Gaussianity with the scale-dependent bias,''
JCAP \textbf{08}, 019 (2012)
[arXiv:1204.6324 [astro-ph.CO]].

\bibitem{Chen:2012ge}
X.~Chen and Y.~Wang,
``Quasi-Single Field Inflation with Large Mass,''
JCAP \textbf{09}, 021 (2012)
[arXiv:1205.0160 [hep-th]].

\bibitem{Pi:2012gf}
S.~Pi and M.~Sasaki,
``Curvature Perturbation Spectrum in Two-field Inflation with a Turning Trajectory,''
JCAP \textbf{10}, 051 (2012)
[arXiv:1205.0161 [hep-th]].

\bibitem{Noumi:2012vr}
T.~Noumi, M.~Yamaguchi and D.~Yokoyama,
``Effective field theory approach to quasi-single field inflation and effects of heavy fields,''
JHEP \textbf{06}, 051 (2013)
[arXiv:1211.1624 [hep-th]].


\bibitem{Cespedes:2013rda}
S.~C\'espedes and G.~A.~Palma,
``Cosmic inflation in a landscape of heavy-fields,''
JCAP \textbf{10}, 051 (2013)
[arXiv:1303.4703 [hep-th]].


\bibitem{Gong:2013sma}
J.~O.~Gong, S.~Pi and M.~Sasaki,
``Equilateral non-Gaussianity from heavy fields,''
JCAP \textbf{11}, 043 (2013)
[arXiv:1306.3691 [hep-th]].


\bibitem{Emami:2013lma}
R.~Emami,
``Spectroscopy of Masses and Couplings during Inflation,''
JCAP \textbf{04}, 031 (2014)
[arXiv:1311.0184 [hep-th]].

\bibitem{Kehagias:2015jha}
A.~Kehagias and A.~Riotto,
``High Energy Physics Signatures from Inflation and Conformal Symmetry of de Sitter,''
Fortsch. Phys. \textbf{63}, 531-542 (2015)
[arXiv:1501.03515 [hep-th]].

\bibitem{Liu:2015tza}
J.~Liu, Y.~Wang and S.~Zhou,
``Inflation with Massive Vector Fields,''
JCAP \textbf{08}, 033 (2015)
[arXiv:1502.05138 [hep-th]].

\bibitem{Arkani-Hamed:2015bza}
N.~Arkani-Hamed and J.~Maldacena,
``Cosmological Collider Physics,''
[arXiv:1503.08043 [hep-th]].

\bibitem{Dimastrogiovanni:2015pla}
E.~Dimastrogiovanni, M.~Fasiello and M.~Kamionkowski,
``Imprints of Massive Primordial Fields on Large-Scale Structure,''
JCAP \textbf{02}, 017 (2016)
[arXiv:1504.05993 [astro-ph.CO]].

\bibitem{Schmidt:2015xka}
F.~Schmidt, N.~E.~Chisari and C.~Dvorkin,
``Imprint of inflation on galaxy shape correlations,''
JCAP \textbf{10}, 032 (2015)
[arXiv:1506.02671 [astro-ph.CO]].

\bibitem{Chen:2015lza}
X.~Chen, M.~H.~Namjoo and Y.~Wang,
``Quantum Primordial Standard Clocks,''
JCAP \textbf{02}, 013 (2016)
[arXiv:1509.03930 [astro-ph.CO]].


\bibitem{Delacretaz:2015edn}
L.~V.~Delacretaz, T.~Noumi and L.~Senatore,
``Boost Breaking in the EFT of Inflation,''
JCAP \textbf{02}, 034 (2017)
[arXiv:1512.04100 [hep-th]].

\bibitem{Bonga:2015urq}
B.~Bonga, S.~Brahma, A.~S.~Deutsch and S.~Shandera,
``Cosmic variance in inflation with two light scalars,''
JCAP \textbf{05}, 018 (2016)
[arXiv:1512.05365 [astro-ph.CO]].

\bibitem{Flauger:2016idt}
R.~Flauger, M.~Mirbabayi, L.~Senatore and E.~Silverstein,
``Productive Interactions: heavy particles and non-Gaussianity,''
JCAP \textbf{10}, 058 (2017)
[arXiv:1606.00513 [hep-th]].

\bibitem{Lee:2016vti}
H.~Lee, D.~Baumann and G.~L.~Pimentel,
``Non-Gaussianity as a Particle Detector,''
JHEP \textbf{12}, 040 (2016)
[arXiv:1607.03735 [hep-th]].


\bibitem{Delacretaz:2016nhw}
L.~V.~Delacretaz, V.~Gorbenko and L.~Senatore,
``The Supersymmetric Effective Field Theory of Inflation,''
JHEP \textbf{03}, 063 (2017)
[arXiv:1610.04227 [hep-th]].

\bibitem{Meerburg:2016zdz}
P.~D.~Meerburg, M.~M\"unchmeyer, J.~B.~Mu\~noz and X.~Chen,
``Prospects for Cosmological Collider Physics,''
JCAP \textbf{03}, 050 (2017)
[arXiv:1610.06559 [astro-ph.CO]].




\bibitem{Chen:2016uwp}
X.~Chen, Y.~Wang and Z.~Z.~Xianyu,
``Standard Model Background of the Cosmological Collider,''
Phys. Rev. Lett. \textbf{118}, no.26, 261302 (2017)
[arXiv:1610.06597 [hep-th]].

\bibitem{Chen:2016hrz}
X.~Chen, Y.~Wang and Z.~Z.~Xianyu,
``Standard Model Mass Spectrum in Inflationary Universe,''
JHEP \textbf{04}, 058 (2017)
[arXiv:1612.08122 [hep-th]].

\bibitem{Kehagias:2017cym}
A.~Kehagias and A.~Riotto,
``On the Inflationary Perturbations of Massive Higher-Spin Fields,''
JCAP \textbf{07}, 046 (2017)
[arXiv:1705.05834 [hep-th]].



\bibitem{An:2017hlx}
H.~An, M.~McAneny, A.~K.~Ridgway and M.~B.~Wise,
``Quasi Single Field Inflation in the non-perturbative regime,''
JHEP \textbf{06}, 105 (2018)
[arXiv:1706.09971 [hep-ph]].

\bibitem{Tong:2017iat}
X.~Tong, Y.~Wang and S.~Zhou,
``On the Effective Field Theory for Quasi-Single Field Inflation,''
JCAP \textbf{11}, 045 (2017)
[arXiv:1708.01709 [astro-ph.CO]].

\bibitem{Iyer:2017qzw}
A.~V.~Iyer, S.~Pi, Y.~Wang, Z.~Wang and S.~Zhou,
``Strongly Coupled Quasi-Single Field Inflation,''
JCAP \textbf{01}, 041 (2018)
[arXiv:1710.03054 [hep-th]].

\bibitem{An:2017rwo}
H.~An, M.~McAneny, A.~K.~Ridgway and M.~B.~Wise,
``Non-Gaussian Enhancements of Galactic Halo Correlations in Quasi-Single Field Inflation,''
Phys. Rev. D \textbf{97}, no.12, 123528 (2018)
[arXiv:1711.02667 [hep-ph]].

\bibitem{Kumar:2017ecc}
S.~Kumar and R.~Sundrum,
``Heavy-Lifting of Gauge Theories By Cosmic Inflation,''
JHEP \textbf{05}, 011 (2018)
[arXiv:1711.03988 [hep-ph]].

\bibitem{Riquelme:2017bxt}
S.~Riquelme M.,
``Non-Gaussianities in a two-field generalization of Natural Inflation,''
JCAP \textbf{04}, 027 (2018)
[arXiv:1711.08549 [astro-ph.CO]].

\bibitem{Franciolini:2017ktv}
G.~Franciolini, A.~Kehagias and A.~Riotto,
``Imprints of Spinning Particles on Primordial Cosmological Perturbations,''
JCAP \textbf{02}, 023 (2018)
[arXiv:1712.06626 [hep-th]].

\bibitem{Saito:2018omt}
R.~Saito and T.~Kubota,
``Heavy Particle Signatures in Cosmological Correlation Functions with Tensor Modes,''
JCAP \textbf{06}, 009 (2018)
[arXiv:1804.06974 [hep-th]].

\bibitem{Cabass:2018roz}
G.~Cabass, E.~Pajer and F.~Schmidt,
``Imprints of Oscillatory Bispectra on Galaxy Clustering,''
JCAP \textbf{09}, 003 (2018)
[arXiv:1804.07295 [astro-ph.CO]].

\bibitem{Wang:2018tbf}
Y.~Wang, Y.~P.~Wu, J.~Yokoyama and S.~Zhou,
``Hybrid Quasi-Single Field Inflation,''
JCAP \textbf{07}, 068 (2018)
[arXiv:1804.07541 [astro-ph.CO]].

\bibitem{Chen:2018xck}
X.~Chen, Y.~Wang and Z.~Z.~Xianyu,
``Neutrino Signatures in Primordial Non-Gaussianities,''
JHEP \textbf{09}, 022 (2018)
[arXiv:1805.02656 [hep-ph]].

\bibitem{Bartolo:2018hjc}
N.~Bartolo, D.~M.~Bianco, R.~Jimenez, S.~Matarrese and L.~Verde,
``Supergravity, $\alpha$-attractors and primordial non-Gaussianity,''
JCAP \textbf{10}, 017 (2018)
[arXiv:1805.04269 [astro-ph.CO]].

\bibitem{Dimastrogiovanni:2018uqy}
E.~Dimastrogiovanni, M.~Fasiello and G.~Tasinato,
``Probing the inflationary particle content: extra spin-2 field,''
JCAP \textbf{08}, 016 (2018)
[arXiv:1806.00850 [astro-ph.CO]].

\bibitem{Bordin:2018pca}
L.~Bordin, P.~Creminelli, A.~Khmelnitsky and L.~Senatore,
``Light Particles with Spin in Inflation,''
JCAP \textbf{10}, 013 (2018)
[arXiv:1806.10587 [hep-th]].

\bibitem{Achucarro:2018ngj}
A.~Ach\'ucarro, S.~C\'espedes, A.~C.~Davis and G.~A.~Palma,
``Constraints on Holographic Multifield Inflation and Models Based on the Hamilton-Jacobi Formalism,''
Phys. Rev. Lett. \textbf{122}, no.19, 191301 (2019)
[arXiv:1809.05341 [hep-th]].


\bibitem{Chua:2018dqh}
W.~Z.~Chua, Q.~Ding, Y.~Wang and S.~Zhou,
``Imprints of Schwinger Effect on Primordial Spectra,''
JHEP \textbf{04}, 066 (2019)
[arXiv:1810.09815 [hep-th]].

\bibitem{Arkani-Hamed:2018kmz}
N.~Arkani-Hamed, D.~Baumann, H.~Lee and G.~L.~Pimentel,
``The Cosmological Bootstrap: Inflationary Correlators from Symmetries and Singularities,''
JHEP \textbf{04}, 105 (2020)
[arXiv:1811.00024 [hep-th]].

\bibitem{Kumar:2018jxz}
S.~Kumar and R.~Sundrum,
``Seeing Higher-Dimensional Grand Unification In Primordial Non-Gaussianities,''
JHEP \textbf{04}, 120 (2019)
[arXiv:1811.11200 [hep-ph]].

\bibitem{Goon:2018fyu}
G.~Goon, K.~Hinterbichler, A.~Joyce and M.~Trodden,
``Shapes of gravity: Tensor non-Gaussianity and massive spin-2 fields,''
JHEP \textbf{10}, 182 (2019)
[arXiv:1812.07571 [hep-th]].

\bibitem{Wu:2018lmx}
Y.~P.~Wu,
``Higgs as heavy-lifted physics during inflation,''
JHEP \textbf{04}, 125 (2019)
[arXiv:1812.10654 [hep-ph]].

\bibitem{Anninos:2019nib}
D.~Anninos, V.~De Luca, G.~Franciolini, A.~Kehagias and A.~Riotto,
``Cosmological Shapes of Higher-Spin Gravity,''
JCAP \textbf{04}, 045 (2019)
[arXiv:1902.01251 [hep-th]].

\bibitem{Li:2019ves}
L.~Li, T.~Nakama, C.~M.~Sou, Y.~Wang and S.~Zhou,
``Gravitational Production of Superheavy Dark Matter and Associated Cosmological Signatures,''
JHEP \textbf{07}, 067 (2019)
[arXiv:1903.08842 [astro-ph.CO]].

\bibitem{McAneny:2019epy}
M.~McAneny and A.~K.~Ridgway,
``New Shapes of Primordial Non-Gaussianity from Quasi-Single Field Inflation with Multiple Isocurvatons,''
Phys. Rev. D \textbf{100}, no.4, 043534 (2019)
[arXiv:1903.11607 [astro-ph.CO]].

\bibitem{Kim:2019wjo}
S.~Kim, T.~Noumi, K.~Takeuchi and S.~Zhou,
``Heavy Spinning Particles from Signs of Primordial Non-Gaussianities: Beyond the Positivity Bounds,''
JHEP \textbf{12}, 107 (2019)
[arXiv:1906.11840 [hep-th]].

\bibitem{Lu:2019tjj}
S.~Lu, Y.~Wang and Z.~Z.~Xianyu,
``A Cosmological Higgs Collider,''
JHEP \textbf{02}, 011 (2020)
[arXiv:1907.07390 [hep-th]].


\bibitem{Hook:2019zxa}
A.~Hook, J.~Huang and D.~Racco,
``Searches for other vacua. Part II. A new Higgstory at the cosmological collider,''
JHEP \textbf{01}, 105 (2020)
[arXiv:1907.10624 [hep-ph]].

\bibitem{Hook:2019vcn}
A.~Hook, J.~Huang and D.~Racco,
``Minimal signatures of the Standard Model in non-Gaussianities,''
Phys. Rev. D \textbf{101}, no.2, 023519 (2020)
[arXiv:1908.00019 [hep-ph]].

\bibitem{Kumar:2019ebj}
S.~Kumar and R.~Sundrum,
``Cosmological Collider Physics and the Curvaton,''
JHEP \textbf{04}, 077 (2020)
[arXiv:1908.11378 [hep-ph]].

\bibitem{Liu:2019fag}
T.~Liu, X.~Tong, Y.~Wang and Z.~Z.~Xianyu,
``Probing P and CP Violations on the Cosmological Collider,''
JHEP \textbf{04}, 189 (2020)
[arXiv:1909.01819 [hep-ph]].


\bibitem{Wang:2019gbi}
L.~T.~Wang and Z.~Z.~Xianyu,
``In Search of Large Signals at the Cosmological Collider,''
JHEP \textbf{02}, 044 (2020)
[arXiv:1910.12876 [hep-ph]].


\bibitem{Wang:2020uic}
Y.~Wang and Y.~Zhu,
``Cosmological Collider Signatures of Massive Vectors from Non-Gaussian Gravitational Waves,''
JCAP \textbf{04}, 049 (2020)
[arXiv:2001.03879 [astro-ph.CO]].

\bibitem{Li:2020xwr}
L.~Li, S.~Lu, Y.~Wang and S.~Zhou,
``Cosmological Signatures of Superheavy Dark Matter,''
JHEP \textbf{07}, 231 (2020)
[arXiv:2002.01131 [hep-ph]].




\bibitem{Wang:2020ioa}
L.~T.~Wang and Z.~Z.~Xianyu,
``Gauge Boson Signals at the Cosmological Collider,''
[arXiv:2004.02887 [hep-ph]].

\bibitem{Bodas:2020yho}
A.~Bodas, S.~Kumar and R.~Sundrum,
``The Scalar Chemical Potential in Cosmological Collider Physics,''
[arXiv:2010.04727 [hep-ph]].



\bibitem{Chen:2016nrs}
X.~Chen, Y.~Wang and Z.~Z.~Xianyu,
``Loop Corrections to Standard Model Fields in Inflation,''
JHEP \textbf{08}, 051 (2016)
[arXiv:1604.07841 [hep-th]].




\bibitem{Stewart:1994ts}
E.~D.~Stewart,
``Inflation, supergravity and superstrings,''
Phys. Rev. D \textbf{51}, 6847-6853 (1995)
[arXiv:hep-ph/9405389 [hep-ph]].



\bibitem{Pinol:2020kvw}
L.~Pinol,
``Multifield inflation beyond $N_\mathrm{field}=2$: non-Gaussianities and single-field effective theory,''
[arXiv:2011.05930 [astro-ph.CO]].

\bibitem{Kallosh:2013yoa}
R.~Kallosh, A.~Linde and D.~Roest,
``Superconformal Inflationary $\alpha$-Attractors,''
JHEP \textbf{11}, 198 (2013)
[arXiv:1311.0472 [hep-th]].



\bibitem{Arnowitt:1962hi}
R.~L.~Arnowitt, S.~Deser and C.~W.~Misner,
``The Dynamics of general relativity,''
Gen. Rel. Grav. \textbf{40}, 1997-2027 (2008)
[arXiv:gr-qc/0405109 [gr-qc]].

\bibitem{Wang:2013zva}
Y.~Wang,
``Inflation, Cosmic Perturbations and Non-Gaussianities,''
Commun. Theor. Phys. \textbf{62}, 109-166 (2014)
[arXiv:1303.1523 [hep-th]].


\bibitem{Maldacena:2002vr}
J.~M.~Maldacena,
``Non-Gaussian features of primordial fluctuations in single field inflationary models,''
JHEP \textbf{05}, 013 (2003)
[arXiv:astro-ph/0210603 [astro-ph]].


\bibitem{Seery:2006vu}
D.~Seery, J.~E.~Lidsey and M.~S.~Sloth,
``The inflationary trispectrum,''
JCAP \textbf{01}, 027 (2007)
[arXiv:astro-ph/0610210 [astro-ph]].

\bibitem{Arroja:2008ga}
F.~Arroja and K.~Koyama,
``Non-gaussianity from the trispectrum in general single field inflation,''
Phys. Rev. D \textbf{77}, 083517 (2008)
[arXiv:0802.1167 [hep-th]].



\bibitem{Chen:2017ryl}
X.~Chen, Y.~Wang and Z.~Z.~Xianyu,
``Schwinger-Keldysh Diagrammatics for Primordial Perturbations,''
JCAP \textbf{12}, 006 (2017)
[arXiv:1703.10166 [hep-th]].




 \end{thebibliography}
\end{document}